\documentclass[amsfonts, amssymb, amsmath, subscriptaddress, showpacs,reprint]{revtex4-1}
\usepackage{times}
\usepackage{bm}
\usepackage{hyperref}
\usepackage{graphicx}
\usepackage{epsfig}

\def\kk{\mathbf k}
\def\pp{\mathbf p}
\def\qq{\mathbf q}
\def\si{{\scriptstyle{K}}}
\def\sii{{\scriptstyle{P}}}
\def\siii{{\scriptstyle{Q}}}
\def\kp{p}
\def\kc{c}
\def\FA{{FS}}

\begin{document}

\title{The energetic coupling of scales in gyrokinetic plasma turbulence}
\author{Bogdan Teaca}
\email{bogdan.teaca@coventry.ac.uk}
\affiliation{Applied Mathematics Research Centre, Coventry University, Coventry CV1 5FB, United Kingdom}
\affiliation{Max-Planck f\"ur Sonnensystemforschung, Justus-von-Liebig-Weg 3, D-37077 G\"ottingen, Germany}
\affiliation{Max-Planck-Institut f\"ur Plasmaphysik, Boltzmannstr. 2, D-85748 Garching, Germany}
\affiliation{Max-Planck/Princeton Center for Plasma Physics}
\author{Alejandro \surname{Ba\~n\'on Navarro}}
\email{alejandro.banon.navarro@ipp.mpg.de}
\affiliation{Max-Planck-Institut f\"ur Plasmaphysik, Boltzmannstr. 2, D-85748 Garching, Germany}
\author{Frank Jenko}  
\email{frank.jenko@ipp.mpg.de}
\affiliation{Max-Planck-Institut f\"ur Plasmaphysik, Boltzmannstr. 2, D-85748 Garching, Germany}
\affiliation{Max-Planck/Princeton Center for Plasma Physics}
\begin{abstract}
In magnetized plasma turbulence, the couplings of perpendicular spatial scales that arise due to the nonlinear interactions are analyzed from the perspective of the free-energy exchanges. The plasmas considered here, with appropriate ion or electron adiabatic electro-neutrality responses, are described by the gyrokinetic formalism in a toroidal magnetic geometry. Turbulence develops due to the electrostatic fluctuations driven by temperature gradient instabilities, either ion temperature gradient (ITG) or electron temperature gradient (ETG). The analysis consists in decomposing the system into a series of scale structures, while accounting separately for contributions made by modes possessing special symmetries (e.g., the zonal flow modes). The interaction of these scales is analyzed using the energy transfer functions, including a forward and backward decomposition, scale fluxes and locality functions. The comparison between the ITG and ETG cases shows that ETG turbulence has a more pronounced classical turbulent behavior, exhibiting a stronger energy cascade, with implications for gyrokinetic turbulence modeling.
\end{abstract}
%
%
\maketitle
%

\section{Introduction}

In many physical systems, nonlinear interactions give rise to couplings between different dynamical scales. In the case of an electrically neutral fluid flow, the velocity field represents the dynamical quantity of interest and the couplings occur between different scales of motion. A similar picture exists for an electrically conductive fluid, where the interplay between velocity and the self-consistent magnetic field gives rise to Alfv\'en waves. The nonlinear interaction can now be interpreted either as the couplings of velocity and magnetic scales or scattering of counter-propagating  Alfv\'en waves of different sizes. For kinetic systems, where the dynamical quantity is represented by a probability distribution function, defined in a six-dimensional phase space, the picture becomes more complicated. Not only that spatial and velocity structures evolve differently, the dynamics can involve moments of the distribution function. These moments can be seen, to an extent, as preferential modes of the distribution function that mediate the nonlinear interactions. 

This situation is present, as well, for magnetized plasmas described kinetically using the gyrokinetic approximation. \cite{Krommes:2012p1373} The gyrokinetic (GK) formalism, valid for a plasma evolving under the influence of a strong magnetic guide field and which obeys the gyrokinetic ordering, represents a self-consistent method of removing the fast gyration phase from the motion of charged particles and decreasing the distribution function phase space from six to five dimensions\cite{Brizard:2007p11}. The influence of the magnetic guide field on the system leads to the development of a spatial anisotropy. Moreover, a similar anisotropy is also developed in velocity space. It should be understood that these anisotropic directions occur at the dynamical level of the equations and that the perpendicular velocity symmetry is taken into account explicitly by the gyrokinetic formalism. Thus, any non-trivial geometry of the magnetic field (non-slab, or in general possessing a non-diagonal metric) will intrinsically complicate the development and saturation of GK turbulence, i.e., the couplings between different dynamical scales. 

For a turbulent GK state, the dynamics determine the type of structures that develop along each direction and influence the linear and nonlinear redistribution of energy in the system. As such, investigating the energy redistribution problem for a spectral form of the GK equations can lead to insight regarding the dynamics of the equations. While different approaches exist for the study of the dynamics introduced by the linear terms, the nonlinear terms drastically limit the available possibilities. To understand the dynamics introduced by the nonlinear term, the redistribution of free-energy (a GK ideal invariant, i.e., a global quantity that remains constant in time in the absence of source and sink effects) is usually investigated. In the current work, we will only look at the redistribution of energy between perpendicular spatial scales, obtained as the integration of contributions along all other directions. While this represents a point of interest in the study of GK turbulence, it should always be remembered that the fundamental dynamics occur in a higher dimensional space that links spatial and velocity dynamics (linear\cite{Hammett:1992p1538, Watanabe:2006p1444} and nonlinear \cite{Schekochihin:2008p1034, Tatsuno:2009p1096, Plunk:2010p1360} phase space mixing), as well as the perpendicular and parallel spatial scales (the concept of critical balance \cite{Barnes:2011p1372}). This is important, as investigating the energetic exchanges of the perpendicular spatial scales only captures part of the nonlinear dynamics.

Understanding the behavior of GK turbulence in the perpendicular directions is important for (heat and particles) transport studies, with relevance to the field of fusion research. The choice of this study to concentrate on the interaction of scales and not of modes\cite{Nakata:2012p1387} is given in part by practical concerns. Promising modeling techniques, like application of Large Eddy Simulations (LES) to GK turbulence \cite{Morel:2011p1339, Morel:2012p1390, BanonNavarro:2014p1535}, rely on the concept of scales, on their separation and on their local energetic interaction. Furthermore, these concepts form the basis of our phenomenological understanding of the problem. To facilitate the analysis, a Fourier mode decomposition of the distribution function in the perpendicular spatial directions is used. In general, well-suited decompositions are needed to give modes proper physical significance (e.g., using spherical harmonics for spherical symmetric problems). Since a natural mode decomposition choice that would depend on the complex geometry is not known to us, we prefer to build our understanding on the phenomenological relevant concept of scales and not on the plane-wave modes. Although the spectral modes enter in the nonlinear dynamics, it is the scales that posses phenomenological information and that can be measured experimentally. The presence of a complex magnetic geometry complicates further the link between dynamical relevant modes and the spatial scales, as perpendicular wavenumbers contribute to more than one perpendicular scale. From this perspective, contributions to perpendicular scales made by modes that possess special symmetries, like the zonal flow, are accounted for separately. While the separation is done at the mode level, the energetic interaction is still performed from the perspective of the scales. This aspect of the problem will be detailed in the current work.

We apply this analysis to ion temperature gradient (ITG) and electron temperature gradient (ETG) turbulence. The mathematical difference between these two cases consists in the adiabatic response given by the electro-neutrality condition, which enters in the GK Poisson equation. The two cases differ by an additional nonlinear term mediated by the electrostatic potential resulting from the flux-surface average contribution to the electro-neutrality condition. As such, in our analysis, we separate the flux-surface average contribution (responsible for the generation of zonal flows) of the electrostatic potential that appears for the ITG case compared to the ETG one. 

Considering the phenomenological interpretation of turbulence, as cascades of energy between scales, decomposing the net transfers into the forward (positive) and backward (negative) contributions allows us to better understand the nature of the energetic exchanges that take place. This analysis is particularly important for the development of more advanced LES models for GK turbulence. Furthermore, looking at the problem from the perspective of scale fluxes, including the scale locality of flux contributions \cite{Kraichnan:1959p497, Teaca:2011p1362} (measuring the contribution of an energy flux through a scale from scales progressively dissimilar in size), provides overall understanding of the GK turbulence problem. To ease the reading of the manuscript, we first introduce the GK equations in Section~\ref{secII}, presenting the difference between the ITG and ETG adiabatic responses and the impact made on the nonlinear term. We continue by presenting the scale decomposition in Section~\ref{secIII}, the transfers analysis in Section~\ref{secIV}, the scale flux analysis in Section~\ref{secV} and end with a discussion of the problem from a theoretical and modeling perspective.

\section{Overview of the problem} \label{secII}

\subsection{The Fourier representation}

For a magnetized plasma, the strong guide field (${\bf B}$) creates an anisotropy in the spatial and velocity directions of the flow. As such, for the study of this system, the use of field-aligned coordinates~\cite{Beer:1995p1292} and the gyrokinetic formalism~\cite{Brizard:1989p1227} represents a natural approach. The field-aligned ${\{x,y,z\}}$ non-orthogonal coordinates parametrize the real space, where $z$ is the coordinate along the magnetic field line (${\bf e}_z$) and the radial coordinate $x$ and the toroidal coordinate $y$ are orthogonal to the magnetic field. The velocity coordinates ${ \{v_{\parallel}, \mu \} }$ are, respectively, the velocity parallel to the magnetic field and the magnetic moment (containing the perpendicular velocity information). Since the turbulent structures tend to align with the magnetic field lines and become elongated along the guide field direction, the definition of the spatial scales becomes entangled with that of the equilibrium magnetic geometry. By using the field-aligned coordinate system to parametrize the real space and by using the same coordinate basis, a Fourier representation can be obtained in the perpendicular direction: $\{x,y\}\rightarrow\{k_x,k_y\}$. The norm of a wave-vector ($\kk=k_x\nabla x+k_y \nabla y$) can be identified with the inverse of a perpendicular spatial scale $k\sim1/\ell$. As the basis used is non-orthogonal, the norm $k\equiv|\kk|$ is defined along the wave-vector $\kk$ direction, using the inner-product $k=[\eta^{ij}k_ik_j]^{1/2}$, were $\eta^{ij}$ are the contra-variant metric tensor components and $i$ and $j$ stand-in for the $\{x,y\}$ components indices. 

In the gyrokinetic representation, the perturbed distribution functions $g_{s} = g_{s}(x, y, z, v_\parallel, \mu, t)$ are the dynamical quantities of interest, where $s$ indices the plasma species and is omitted altogether when the adiabatic approximation \cite{Hammett:1993p1536} is used for the electro-neutrality response. Each perturbed distribution function, characterising a species of particles of charge $q$ and mass $m$, is assumed to evolve around an equilibrium point given by an appropriately normalised\cite{Merz2009} Maxwellian contribution $F_{0}=\pi^{-3/2}e^{-(v_\parallel^2+\mu B_0)}$ and which possess a background global temperature $T_{0}$. This allows for the decomposition $g=h - q \frac{F_0}{T_0}\phi$ to be made, where $h$ is the non-adiabatic part of the perturbed distribution functions and $\phi=\phi[g]$ is the gyro-averaged self-consistent electrostatic field contribution.

\subsection{The gyrokinetic equations}

The gyrokinetic equations are solved using the Eulerian code {\sc Gene}\cite{Jenko:2000p1248}, which can be used in both global and local (flux-tube) approximation.\cite{Gorler:2011p1340} For simplicity, we restrict the problem to the local approximation of a toroidal magnetic equilibrium configuration, using the electrostatic approximation (no magnetic fluctuations). Symbolically, the gyrokinetic evolution equations can be expressed as
\begin{align}
\frac{ \partial  g} {\partial t} =  G[g] + L [g] + D[g] + N[g,g]  \;. \label{eq_GK}
\end{align}
With the exception of the last term, all others are linear in $g$. The functional dependence is taken as $[g]$, even for terms that depend explicitly on $\phi$ and $h$, as the electric field and the non-adiabatic contribution to the distribution function can be determined from $g$. The first term in Eq.~(\ref{eq_GK}) is due to the contribution of the normalised background density ($\omega_{n}$) and temperature ($\omega_{T}$) gradients acting on the electrostatic field gradients and represents the driving mechanism for GK turbulence, 
\begin{align}
G[g]=&-\left[\omega_{n}+\left( v_\parallel^2+\mu B_0-\frac{3}{2}\right) \omega_{T}\right] F_{0}\frac{\partial \phi}{\partial y} \label{eq_GK1}\;.
\end{align}
The second linear term appears due to the magnetic curvature ($K_x$ and $K_x$ factors depend on the magnetic geometry and the respective metric coefficients) and contains the parallel dynamics involving magnetic trapping and linear Landau damping/pumping effects,
\begin{align}
L[g]=&-\frac{T_{0}(2v_\parallel^2+\mu B_0)}{q B_0}\left( K_x\frac{\partial h}{\partial x}+K_y\frac{\partial h}{\partial y} \right) \nonumber\\
&-\frac{v_{T}}{2}\left[\mu \frac{ B_0 }{\partial z} \frac{\partial h}{\partial v_\parallel}  -  \frac{\partial  (v_\parallel^2+\mu B_0) }{\partial v_\parallel} \frac{\partial h}{\partial z} \right]\;, \label{eq_GK2}
\end{align}
where $v_{T}=\sqrt{2T_{0}/m}$ is the thermal velocity. The third linear term contains the dissipative effects. The dissipation terms have a simple hyper-diffusivity form, 
\begin{align}
D[g]=&-\bigg{(}a_z \frac{\partial^n }{\partial z^n} + a_{v_{\parallel}} \frac{\partial^n }{\partial v_{\parallel}^n} \bigg{)} g\;, \label{eq_GK3}
\end{align}
where $n=4$ and the $a$'s parameters are adapted to the problem at hand. A collision operator can also be included.

Finally, the last term contains the $\mathbf E\times \mathbf B$ drift nonlinearity. This term has the fundamental role of coupling different five dimensional scales in phase space and leads to an effective coupling of perpendicular scale structures,
\begin{align}
N[g,g]=& \frac{\partial \phi}{\partial y}\frac{\partial h}{\partial x} - \frac{\partial \phi}{\partial x}\frac{\partial h}{\partial y}\;.
\label{nonlin}
\end{align}
While all terms contribute to the balance equation, it is on the nonlinear term that we will concentrate our analysis. Moreover, as the metric $\eta^{ij}$ depends on $z$ for most magnetic geometries of interest, the same $k_x, k_y$ wavenumbers can contribute to different scale lengths ($\ell_k$). This is important as the nonlinear interaction term is defined in terms of $k_x, k_y$ wavenumber interactions. 

Indeed, omitting the velocity dependences, which are not of immediate interest, we see that the nonlinear term is given by a Poisson bracket structure in the $\{x,y\}$ space which for the $\{k_x,k_y\}$ Fourier representation simply becomes,
\begin{align}
N(k_x,\!k_y,\! z)= \!\!\!\!\!\!\!\!\!\!\!\sum_{\substack{k_x-p_x-q_x=0\\ k_y-p_y-q_y=0}} \!\!\!\!\!\!\!\!\![q_xp_y\!-\!q_yp_x]\, \phi(q_x,\!q_y,\! z)h(p_x,\!p_y,\!z) \label{eq_NL}\;. 
\end{align}
From this form, it is clear that specifying the $k_x-p_x-q_x=0$ and $k_y-p_y-q_y=0$ wavenumber vertex interactions will not define the nonlinear coupling of scales denoted by the norms $k\sim 1/\ell_k$, $p\sim 1/\ell_p$ and $q\sim 1/\ell_q$, due to the $z$ dependence of the metric tensor that enters in the definition of the scales.

\subsection{The quasi-neutrality responses}

We see that the GK equation (Eqs.~\ref{eq_GK}-\ref{nonlin}) has the same form regardless of the species studied, be it ions or electrons. However, to obtain a closed system, the self-consistent gyro-averaged electrostatic potential ($\phi$) needs to be obtained. In the Fourier representation, the gyro-averaged electrostatic potential is simply the Bessel function ($J_0$) screened electrostatic potential ($\varphi$), i.e. $\phi(\kk, z)=J_0(\lambda) \varphi(\kk, z)$, with $\lambda=\sqrt{\mu B_0}k v_{T}/\Omega$ and $\Omega=qB_0/(m \mbox{c})$. 

\begin{figure}[tb]
\begin{center}
\includegraphics[width = 0.43\textwidth]{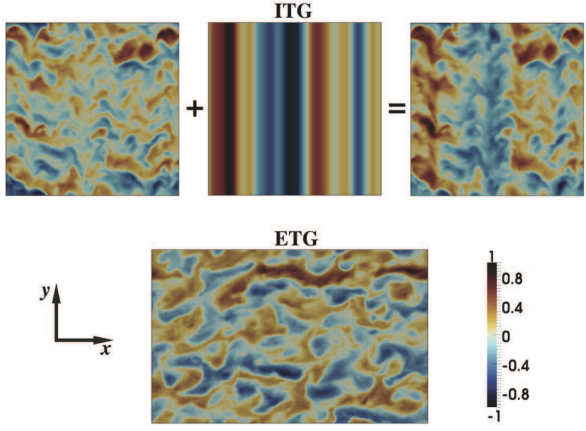}
\end{center}
\caption{(Color online) Cross-section through the gyro-averaged electrostatic potentials for ITG and ETG cases. For ITG, the contributions made by the terms given in Eq.~(\ref{eq_ITG}) are also presented. Details of the simulations are given in sub-section~\ref{secNS}.}
\label{fig_phi}
\end{figure}

Using the adiabatic approximation for one of the species (of density $\delta n$), the electrostatic potential [$\varphi(\kk, z)$] is found from the simplified gyrokinetic Poisson equation
\begin{align}
\frac{\delta n}{n_0}&=\pi B_0\int J_0(\lambda) g\; \mbox{d}v_{\parallel} \mbox{d}\mu + [1-\Gamma_0(b)] \frac{q\varphi}{T_0}\;
\end{align}
and requires knowledge of the (charge) density fluctuations of all plasma constituents (including $\delta n$). It is at this stage that the difference between the ITG and ETG cases becomes apparent, in the form of the adiabatic response (${\delta n}/{n_0}$) considered for the adiabatic species, 
\begin{align}
\frac{\delta n}{n_0}&=\frac{q\varphi}{T_0},\ \ \ \mbox{for ETG, and} \; \\
\frac{\delta n}{n_0}&=\frac{q(\varphi-\langle\varphi\rangle_{FS})}{T_0},\ \ \  \mbox{for ITG}\;.
\end{align}
The ion response incorporates the departure from the flux-surface average ($\langle\cdots\rangle_{FS}$ defined below in Eq.~\ref{FSaverage}), which leads to an additional contribution. While in the ETG case (considered for electrons) the electrostatic potential is found as
\begin{align}
\varphi^{\sc ETG}=\frac{1}{1-\Gamma_0(b)+\tau} \frac{\pi B_0 T_0}{q}\!\! \int \!\!J_0(\lambda) g \mbox{d}v_{\parallel} \mbox{d}\mu\; ,\label{eq_ETG}
\end{align}
for the ITG case (considered for ions) the electrostatic potential takes the form
\begin{align}
\varphi^{\sc ITG}&=\frac{1}{1-\Gamma_0(b)+\tau}\Big{[}\frac{\pi B_0 T_0}{q}\!\! \int \!\! J_0(\lambda) g \mbox{d}v_{\parallel} \mbox{d}\mu+\tau\langle\varphi\rangle_{FS}\Big{]} \nonumber \\
&=\varphi^{\sc ETG} + \frac{\tau\langle\varphi^{\sc ITG}\rangle_{FS}}{1-\Gamma_0(b)+\tau} \label{eq_ITG}.
\end{align}
As example, the contributions made by the two terms to the ITG and ETG gyro-average potentials are presented in FIG.~\ref{fig_phi}.

Performing a $\delta\varphi=\varphi-\langle\varphi\rangle_{FS}$ decomposition, we see that the $\delta\varphi$ part of the potential is identical in the two cases. The different form of the adiabatic response just changes the flux-surface averaged contribution,
\begin{align}
\langle\varphi^{\sc ETG}\rangle_{FS}&= \frac{1}{1-\Gamma_0(b)+\tau}\Big{<}\frac{\pi B_0 T_0}{q}\!\! \int \!\! J_0(\lambda) g \mbox{d}v_{\parallel} \mbox{d}\mu\Big{>}_{FS}\;, \label{fsITG}  \\
\langle\varphi^{\sc ITG}\rangle_{FS}&=\frac{1}{1-\Gamma_0(b)}\Big{<}\frac{\pi B_0 T_0}{q}\!\! \int \!\! J_0(\lambda) g \mbox{d}v_{\parallel} \mbox{d}\mu\Big{>}_{FS}\;, \label{fsETG}
\end{align}
resulting in a less damped contribution in the ITG case. In all of above, the function $\Gamma_0(b)=e^{b}I_0(b)$ and the modified Bessel function $I_0$ have the arguments defined as $b=k^2 v^2_{T}/(2\Omega^2)$, while $\tau$ represents the kinetic to adiabatic species temperature ratios (ion to electron temperature ratio for ITG case and the electron to ion temperature ratio for the ETG case).

\subsection{A note on the nonlinear term}

Taking into account the ITG and ETG form of the gyro-averaged electrostatic potential ($\phi$) (resulting from Eq.~\ref{eq_ETG} and Eq.~\ref{eq_ITG}) entering in the nonlinear term, we see that the ITG nonlinearity differs by an additional flux-surface average contribution (it is also interesting to note that this is the only place where this additional contribution enters in the GK evolution equations). This contribution changes the ${\langle\phi^{\sc ITG}\rangle_{FS}}$ and ${\langle\phi^{\sc ETG}\rangle_{FS}}$ signal entering into the nonlinear term. 
\begin{align}
N^{\sc ITG}&=N^{\delta \phi}+N^{\langle\varphi^{\sc ITG}\rangle_{FS}}  \;,\\
N^{\sc ETG}&= N^{\delta \phi}+N^{\langle\varphi^{\sc ETG}\rangle_{FS}} \;.
\end{align}

Although symbolically we can extract a common part of the nonlinear interactions, due to the intrinsic nonlinear character of the problem, we cannot extract a common flow behavior corresponding to only one term. The stronger interactions for ITG ($N^{\langle\varphi^{\sc ITG}\rangle_{FS}} >N^{\langle\varphi^{\sc ETG}\rangle_{FS}}$, since $\langle\varphi^{\sc ITG}\rangle_{FS}>\langle\varphi^{\sc ETG}\rangle_{FS}$), will change the overall behavior of $N^{\delta \phi}$ in the two cases.

\subsection {Numerical parameters} \label{secNS}

The geometry used is a concentric circular model \cite{Lapillonne:2009p1355} and both ITG and ETG cases use a resolution of $256 \times 256 \times 24 \times 48 \times 16$ in the $x \times y \times z \times v_{\parallel} \times \mu$ coordinates. However, the perpendicular box sizes $L_x \times L_y$ are different in the two cases. For ITG a box size of $125\, \rho\times 125 \, \rho$ is used, while for ETG we employ a box of $200\, \rho \times 125\, \rho$ to account for streamers in the $x$ direction, where $\rho$ represents the gyroradius of the kinetic species. For ITG, the typical CBC parameters \cite{Dimits:2000p1375} are used:  safety factor $q=1.4$, magnetic shear $\hat{s} = 0.8$, aspect ratio (small radius $r$ over the large radius $R$) of  $r/R = 0.18$, temperature and density gradients of  $\omega_{T} = 7$ and $\omega_n = 2.2$.  In addition, for the ITG case we employ a Landau-Boltzmann collision operator with collision frequency of $\nu = 0.005 [v_{T}/R]$ and hyper-diffusion in $z$ and $v_{\parallel}$, with coefficients of $a_z = 1.0$ and $a_v=0.2$.

For the ETG case, parameters are taken as the CBC ones as well (listed above for ITG), but with a lower safety factor of $\hat{s} = 0.1$. This choice is made to be in line with the authors previous LES work\cite{BanonNavarro:2014p1535} and ETG benchmarking efforts\cite{Nevins:2006p1454}. Choosing a lower magnetic shear is known to reduce the heat transport for ETG turbulence\cite{Jenko:2000p1248, Dorland:2000p1088}. However, for ETG no collision operator is used, as physically electron-electron collisions are not expected to be sufficiently strong. Hyper-diffusion in $z$ and $v_{\parallel}$ with the same coefficients as for the ITG case is used.

\section{The energetic picture} \label{secIII}

The free energy ($\mathcal{E}$) represents the quadratic quantity of interest for the study of gyrokinetic turbulent dynamics \cite{Schekochihin:2009p1131}. Free-energy is the quantity that is injected into the system by the gradients and dissipated by collisions, while being redistributed in a conservative fashion by the action of the nonlinear term. Formally, the free energy is defined as
\begin{align}
\mathcal{E}  = \left<\frac{T_0}{2F_0}h g \right>_{\Lambda}\;,
\end{align}
where $\Lambda$ is the phase space volume (the volume element being $\mbox{d}\Lambda= \sqrt{\eta} (\pi B_0 n_0) \mbox{d} k_x  \mbox{d}k_y\mbox{d}z\mbox{d} v_{\parallel}\mbox{d}\mu$, with $\sqrt{\eta}=\det [\eta^{ij}]^{-1/2}$ the field-aligned Jacobian) and the $\left<\cdots \right>_X$ notation stands for the average over the $X$ domain; see Ref.~[\onlinecite{ABNThesis}] for a full working of free energy definition. Considering our interest in analyzing the energetic coupling of perpendicular scales, first we look at the balance equation for a mode before concentrating on scales interaction.

\subsection{The free-energy balance equation for a mode}

Working in the Fourier representation for the perpendicular spatial directions, we can define the spectral density of free energy for a mode identified by the wave-vector $\kk$ as
\begin{align}
\mathcal{E}(\kk)  = \left<\frac{T_0}{2F_0}h(-\kk) g(\kk) \right>_\Theta, 
\end{align}
where $\Theta$ is implicitly defined from the phase space volume element ($\mbox{d}\Lambda= \mbox{d}k_x \mbox{d}k_y\mbox{d}\Theta$) and considering the reality condition with respect to Hermitian conjugation $h^*(\kk)=h(-\kk)$. From the GK equations (Eq.~\ref{eq_GK}), the free-energy balance equation for a mode can be written as
\begin{align}
\frac{\partial  \mathcal{E}(\kk) } {\partial t}  = \mathcal{G}(\kk) + \mathcal{L}(\kk)  + \mathcal{D}(\kk)+  \mathcal{T}(\kk)\;.
\label{eq:FE_spectral}
\end{align}
where the linear terms $\mathcal A=\{ \mathcal{G} , \mathcal{L}  , \mathcal{D} \}$ are computed from the terms entering in the GK equation $ A=\{ {G} , {L}  , {D} \}$ as
\begin{align}
 \mathcal{A}   =  \left< \frac{T_0}{2F_0} h(-\kk) A(\kk)\right>_\Theta\;.
\label{eq_ener_shell}
\end{align}
With respect to the terms on the right-hand side of the equation, $\mathcal{G}(\kk)$ represents the free energy injected into the system for a mode $\kk$ by the temperature and density gradients, $\mathcal{L}(\kk)$ is the linear contribution composed by the parallel and curvature terms (and globally integrates to zero). The term  $\mathcal{D}(\kk)$ is the local dissipation and finally, $\mathcal{T}(\kk)$ is the nonlinear free-energy transfer term. 

The free-energy transfer represents the energetic contribution of the nonlinear term (Eq.~\ref{eq_NL}) and has the form,
\begin{align}
\mathcal T(\kk)=  \left<\frac{T_0}{2F_0} h(-\kk)  N(\kk) \right>_\Theta=\sum_{\pp, \qq} \mathcal T({\mathbf k| \mathbf p, \mathbf q}),
\end{align}
where $\mathcal T({\mathbf k| \mathbf p, \mathbf q})$ is the triad transfer representing the redistribution of free energy between modes $\kk$ due to the interaction with modes $\pp$ and $\qq$, if $\kk+\pp+\qq={\bf 0}$ and is zero otherwise. Since the perpendicular-scale triad transfer is integrated over all other directions, it represents an effective mechanism for the exchange of free energy. It should not be seen as the fundamental energetic interaction for gyrokinetics, as such object is defined in the full five-dimensional phase space.

\subsection{The free-energy transfers between modes}

Taking into account in a manifest way the $\qq$ and $\pp$ symmetry arising from the convolution, we define the triad transfer as
\begin{align}
\mathcal T(\kk|\pp,\qq) = \bigg{<}\frac{T_0}{4F_{0}}\! &\big{[} q_x p_y\!  - \! q_y  p_x \big{]}  \times \nonumber \\
&\big{[} \phi(\qq) h(\pp)\!  - \! \phi (\pp) h(\qq) \big{]}  h(\kk) \bigg{>}_\Theta \label{eq_triad} 
\end{align}
for $\kk+\pp+\qq={\bf 0}$ and zero otherwise. While the nonlinear term is symmetric in $\qq$ and $\pp$, we see that this symmetry is achieved by the product of two anti-symmetric structures, the Poisson bracket one (resulting in the $\big{[} q_x p_y  - q_y  p_x \big{]}$ geometric contribution) and the field operator corresponding to the Poisson equation that relates $\phi$ to $g$ .

In addition to the $\mathcal T({\mathbf k| \mathbf p, \mathbf q})=\mathcal T({\mathbf k| \mathbf q, \mathbf p})$ symmetry being evident, the energy conservation in a triad can also be easily determined,
\begin{align}
\mathcal T(\mathbf k| \mathbf p, \mathbf q)+\mathcal T(\mathbf p| \mathbf q, \mathbf k)+\mathcal T(\mathbf q| \mathbf k, \mathbf p)=0\;.
\label{triad_conservation}
\end{align}
Although this object correctly accounts for the triad transfer and cumulatively accounts for all energetic fluxed quantities that arise, it is more convenient to split the triad transfer into mode-to-mode transfer,
\begin{align}
 \mathcal T(\kk|\pp,\qq) &= \Big{[} \mathcal S(\kk|\pp|\qq) +\mathcal S(\kk|\qq|\pp)\Big{]} , 
\end{align}
where
\begin{align}
\mathcal S(\kk|\pp|\qq)&= \bigg{<}\frac{T_0}{4F_{0}}\big{[} q_x p_y\!  - \! q_y  p_x \big{]} \phi(\qq) h(\pp)  h(\kk) \bigg{>}_\Theta. \label{eq_mode}
\end{align}

For the mode-to-mode transfer, the position of each mode that enters the definition matters. This non-unique decomposition (up to a circulation transfer that sums up to zero in a triad\cite{Verma:2004p206}) allows for a more detailed interpretation of the transfers. For instance, since the anti-symmetry of $\pp$ and $\kk$ is evident, we can interpret $S(\kk|\pp|\qq)$ as the energy that mode $\kk$ receives from mode $\pp$ due to the mediation of mode $\qq$ and is opposite in value to the energy that mode $\pp$ receives from mode $\kk$ due to the mediation of the same mode $\qq$, i.e. $S(\kk|\pp|\qq)=-S(\pp|\kk|\qq)$. This conceptual decomposition is solely done to help with the interpretation.

\subsection{The zonal-flow mode transfers}  
  
The mode-to-mode decomposition allows to identify the contribution made to a turbulent state by special modes, like the modes responsible for the zonal flows. For a GK plasma in toroidal geometry, the zonal flow contribution is given by a flux surface averaged signal. In our representation, the flux surface average is defined as
\begin{align}
&\langle g(k_x,k_y,z)\rangle_{\FA}=\frac{1}{V_{\FA}}\int g(k_x, 0, z) \sqrt{\eta}\ dz \;, \label{FSaverage}
\end{align}
where $V_{\FA}=\iint \sqrt{\eta}\ \mbox{d}y \mbox{d}z$ and we used $\langle g(k_y)\rangle_y=g(0)$. While the flux surface integration over $z$ is crucial, we denote modes $\kk^{ZF}=(k_x, 0)$ as zonal-flow modes. These are the modes that contribute to the zonal-flow signal, across all scales. Depending on the role of the zonal flow mode entering in the mode-to-mode interaction and due to the three wave resonance condition, we have      
\begin{align}
\mathcal S(\kk^{ZF}|\pp|\qq)= \bigg{<}\frac{T_0}{4F_{0}}\!  \Big{[}  &-p_y  k_x \Big{]}\phi(q_x, -p_y, z)\times \nonumber \\ \, 
&h(p_x, p_y, z)\, h(k_x, 0, z)\bigg{>}_\Theta \ , \\ 
\mathcal S(\kk|\pp^{ZF}|\qq)= \bigg{<}\frac{T_0}{4F_{0}}\! \Big{[} &+p_x k_y \Big{]}\phi(q_x, -k_y, z)\times \nonumber \\ \, 
&h(p_x,0,z)\, h(k_x, k_y, z) \bigg{>}_\Theta\ ,\\ 
\mathcal S(\kk|\pp|\qq^{ZF})=\bigg{<} \frac{T_0}{4F_{0}}\! \Big{[}  &-k_y  q_x \Big{]}\phi(q_x,0,z)\times \nonumber \\ \, 
&h(p_x, -k_y, z)\, h(k_x, k_y, z) \bigg{>}_\Theta\ .
\end{align}
The mediation of the zonal flow only contains the $S(\kk|\pp|\qq^{ZF})$ interactions, responsible for the transfer of energy to smaller $k_x$ scales. $S(\kk^{ZF}|\pp|\qq)$ and $S(\kk|\pp^{ZF}|\qq)$ only differ by a minus sign and are responsible for a nonlocal energy movement. We expect these interactions to contribute drastically to the non-locality of energy interactions. This does not come as a surprise, since phenomenologically the shearing of small scales by large scales contributes to nonlocal interactions\cite{Howes:2011p1459} and the zonal flow modes mainly contribute to large scale motions.  

From the perspective of the triad transfer, only two sets of interactions are possible for the zonal flow modes,
\begin{align}
\mathcal T(\kk^{ZF}|\pp,\qq) &= \Big{[}  \mathcal S(\kk^{ZF}|\pp|\qq) +\mathcal S(\kk^{ZF}|\qq|\pp)\Big{]} ,\\ 
\mathcal T(\kk|\pp^{ZF},\qq) &= \Big{[}  \mathcal S(\kk|\pp^{ZF}|\qq) +\mathcal S(\kk|\qq|\pp^{ZF})\Big{]} . 
\end{align}

We note that due to the difference in the definition of the adiabatic response for ITG and ETG, best seen in Eq.~(\ref{eq_ITG}), the contribution of the $\kk^{ZF}$ mode is accentuated for ITG. It does not imply that ETG does not possess this contribution, it just means that the zonal flow modes act stronger for ITG, Eqs.~(\ref{fsITG}-\ref{fsETG}). Phenomenologically, it coincides with an accentuated sheared rotation of the nested flux surfaces for ITG, when compared to ETG and thus, an accentuated shearing of smaller turbulent structures (energy transfers being mediated by the $\kk^{ZF}$ modes).

\subsection{The scale decomposition}

Starting from the idea that scales are the structures of physical importance, not the wave modes, we interpret the non-linear interactions as in the case of classical turbulence, looking at the coupling of scales in the system. For non-orthogonal coordinates, as is the case here, we accept that the same $k_x, k_y$ wavenumber interaction contributes to different scales.  

As in similar works \cite{BanonNavarro:2011p1274, Teaca:2012p1415}, we start from the decomposition of the space into scale structures, identical to selecting structures $s_\si=[k_{\si-1},k_{\si}]$, with boundary wavenumbers given as a geometric progression ($k_\si=k_0\lambda^\si$, here $\lambda= 2^{1/5}$). We call these structures shells due to previous uses in literature, however it should be noted that the geometric shape is not always that of a cylindrical or spherical shells. We mention that while a infinitesimal decomposition could be performed, equivalent to recovering the wave-norm $k$, a geometric progression is preferred for turbulence studies, since scaling laws play an important part.

The nonlinear transfer between shells represents a diagnostic that consists in filtering the distribution function before building the free-energy transfer functions. The shell-filtered distribution functions $g^\si(\kk)$ are found as
\begin{eqnarray}
g^\si(\kk)=\left\{ \begin{array}{lcl}
g(\kk) , &  |\kk| \in s_\si \\
0 , &  |\kk| \notin s_\si  
\end{array}  \right.  \;. 
\end{eqnarray}
It is important to realize that the shell-filtered distribution functions are well defined in real space, the total signal being recovered as the superposition of all scale filtered contributions, 
\begin{align}
g(x,y)=\sum_\si g^\si(x,y)\;.
\end{align}

\subsection{The free-energy interaction for a scale}

\begin{figure}[b]
\begin{center}
\includegraphics[width = 0.35\textwidth]{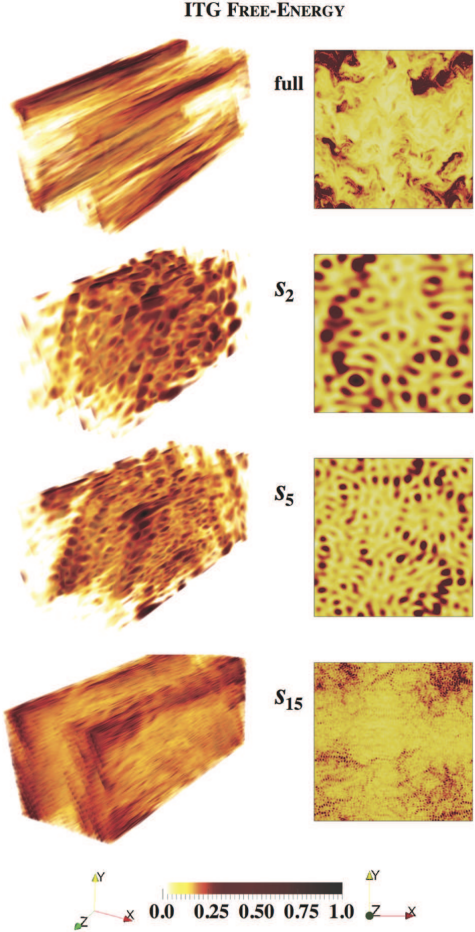}
\end{center}
\caption{(Color online) The free-energy density structures and filtered contributions arising from different shells. The energy density is normalized by its global integrated value. On the right, $z=0$ cross-section planes are shown for clarity.}
\label{fig_scale_separation}
\end{figure}

In relation to the mode structure, the free energy contained in a scale can be easily found as
\begin{align}
\mathcal{E}(\si)  &= \left<\ \int\limits_{|\kk| \in s_\si}\!\!\!\! \mathcal{E}(\kk) \mbox{d}\kk\right>_\Theta \nonumber \\
&= \left<\ \int\limits_{|\kk| \in s_\si}\!\!\!\!\frac{T_0}{2F_0}h(\kk) g(\kk) \mbox{d}\kk \right>_\Theta  \nonumber \\
& = \left< \frac{T_0}{2F_0}h(\kk) g^\si(\kk) \right>_\Lambda= \left< \frac{T_0}{2F_0}h^\si(\kk) g(\kk) \right>_\Lambda, 
\end{align}
emphasizing the importance of the order of the integrals and using the fact that two scales are orthogonal to each other (arising from the definition). As example, for the ITG case, we show in FIG.~\ref{fig_scale_separation} the free-energy contained in three different shells as a representation of scales, as well as the unfiltered quantity. As we can see, ever larger indexed shells contain smaller and smaller structures. We mention that we have 30 shells in total. This allows us to interpret the nonlinear interactions as the coupling between "eddies" of different size. 
 
\begin{figure}[b]
\begin{center}
\includegraphics[width = 0.40\textwidth]{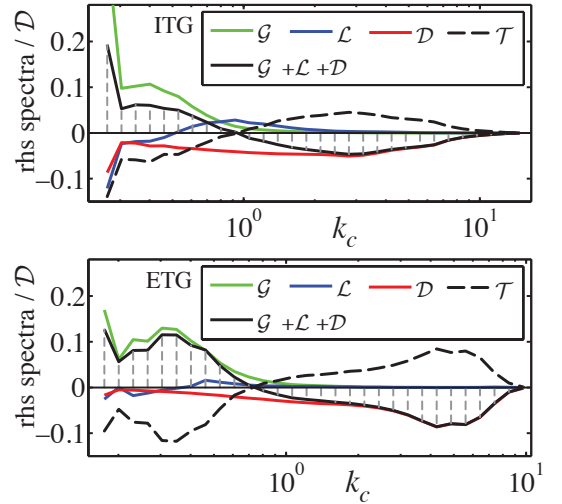}
\end{center}
\caption{(Color online) The free-energy balance equation {\em rhs} spectra, normalized to the total dissipation. The vertical dashed lines represent the shell boundaries ($k_\kc$), with $c$ taking the shell index value (between $1$ and $30$).}
\label{fig_linear}
\end{figure}

The balance equation for a shell can now be found in a similar manner, giving,
\begin{align}
\frac{\partial  \mathcal{E}(\si) } {\partial t}  = \mathcal{G}(\si) + \mathcal{L}(\si)  + \mathcal{D}(\si)+  \mathcal{T}(\si)\;,
\label{eq_ener_shell}
\end{align}
where the {\em rhs} terms $\{ \mathcal{G}(\si) , \mathcal{L}(\si)  , \mathcal{D}(\si) , \mathcal{T}(\si)\}$ are computed in a similar fashion as the energy, as the filtered contribution of their respective mode quantities. For the two cases studied, we plot the linear terms spectra (the {\em rhs} terms of Eq.~\ref{eq_ener_shell}) in FIG.~\ref{fig_linear}. Compared to ITG, the ETG case has smaller dissipation at large scales. For ETG, the dissipation tends to peak at small scales. In both cases, the sum of the linear contributions at each scale balances the nonlinear transfer spectra ($\mathcal T$).

Due to the orthogonality of shells, the contribution arising from the nonlinear term poses no complication in being computed directly. However, the nonlinear term can be computed as the interaction of three scales $\si$, $\sii$ and $\siii$. Starting from the triad-transfer definition (Eq.~\ref{eq_triad}), we filter the fields before computing the transfer,
\begin{align}
\bar{\mathcal T}(\kk|\pp,\qq)& = \bigg{<}\frac{T_0}{4F_{0}}\! \big{[} q_x p_y\!  - \! q_y  p_x \big{]}  \times \nonumber \\
&\big{[} \phi^\siii(\qq) h^\sii(\pp)\!  - \! \phi^\siii (\pp) h^\sii(\qq) \big{]}  h^\si(\kk) \bigg{>}_\Theta. \label{eq_triad_brk} 
\end{align}
For $\bar{ \mathcal T}({\mathbf k| \mathbf p, \mathbf q})$, the manifest symmetry in ${\mathbf q}$ and ${\mathbf p}$ of the triad transfers is broken effectively by the shell filtering procedure, as $h^\sii(\qq)=0$ for $q\notin s_\sii$. We see that the same result can be obtained starting from Eq.~(\ref{eq_mode}). We thus define the triple-scale (shell) transfer as
\begin{align}
{\mathcal S}(\si|\sii|\siii)& \!=\! \bigg{<}\! \frac{T_0}{2F_{0}}\!\ \! \big{[} q_x p_y\!  - \! q_y  p_x \big{]} \phi^\siii\!(\qq) h^\sii\!(\pp)  h^\si\!(\kk) \!\! \bigg{>}_\Lambda, \label{eq_tripleshell}
\end{align}
or equivalently in real space as
\begin{align}
\mathcal  S(\si|\sii|\siii)=& \bigg{<}\frac{T_0}{2F_{0}}\bigg{[}\frac{\partial \phi^\siii}{\partial y}\frac{\partial h^\sii}{\partial x} - \frac{\partial \phi^\siii}{\partial x}\frac{\partial h^\sii}{\partial y}\bigg{]}h^\si\bigg{>}_\Lambda\;.
\end{align}
The real space form of the definition can be more intuitive for global simulation works or non-axisymmetric geometric conditions. We notice that the averaging is done over the entire volume as the scale separation is given by the filtering procedure. 

Similarly, we define the three contributions made by the zonal-flow modes to the triple-scale (shell) transfer as
\begin{align}
{\mathcal S}(\si^{ZF}|\sii|\siii)& \!=\! \bigg{<}\! \frac{T_0}{2F_{0}}\!\ \! \big{[} q_x p_y\!  - \! q_y  p_x \big{]} \phi^\siii\!(\qq) h^\sii\!(\pp)  h^\si\!(\kk^{ZF}) \!\! \bigg{>}_\Lambda, \\
{\mathcal S}(\si|\sii^{ZF}|\siii)& \!=\! \bigg{<}\! \frac{T_0}{2F_{0}}\!\ \! \big{[} q_x p_y\!  - \! q_y  p_x \big{]} \phi^\siii\!(\qq) h^\sii\!(\pp^{ZF})  h^\si\!(\kk) \!\! \bigg{>}_\Lambda, \\
{\mathcal S}(\si|\sii|\siii^{ZF})& \!=\! \bigg{<}\! \frac{T_0}{2F_{0}}\!\ \! \big{[} q_x p_y\!  - \! q_y  p_x \big{]} \phi^\siii\!(\qq^{ZF}) h^\sii\!(\pp)  h^\si\!(\kk) \!\! \bigg{>}_\Lambda, 
\end{align}
or a total contribution
\begin{align}
{\mathcal S}^{ZF}(\si|\sii|\siii)={\mathcal S}(\si^{ZF}|\sii|\siii)+{\mathcal S}(\si|\sii^{ZF}|\siii)+{\mathcal S}(\si|\sii|\siii^{ZF})
\end{align}
that contains the receiver, giver and mediator contributions to a given scale, without differentiating between the three.

\section{Scale-to-scale transfers} \label{secIV}

For GK turbulence, the scale-to-scale (shell-to-shell) transfers have been studied before in the literature\cite{Tatsuno:2010p1363, BanonNavarro:2011p1274}. They represent one of the first type of nonlinear diagnostics to be adopted by the field of plasma turbulence\cite{Camargo:1995p1564} from the field of hydrodynamical (classical) turbulence.\cite{Domaradzki:1990p145} From the triple-scale transfer, they are defined as  
 \begin{align}
{\mathcal P}(\si|\sii) =\sum_{\siii}{\mathcal S}(\si|\sii|\siii)
\label{s2s_def}\;.
\end{align}
It has the interpretation of the energy received by modes located in a shell $\si$ from modes located in a shell $\sii$ by the interaction with all other possible modes. Due to the conservation of interaction, $\mathcal P(\si|\sii) =-\mathcal P(\sii|\si) $ and $\mathcal P(\si|\si) =0$ for each species. Furthermore, it allows the recovery of the non-linear transfer spectra,
 \begin{align}
{\mathcal T}(\si) =\sum_{\sii}{\mathcal P}(\si|\sii) =\sum_{\sii}\sum_{\siii}{\mathcal S}(\si|\sii|\siii)\;.
\label{lt_def}
\end{align}
%

\begin{figure}[tb]
\begin{center}
\includegraphics[width = 0.48\textwidth]{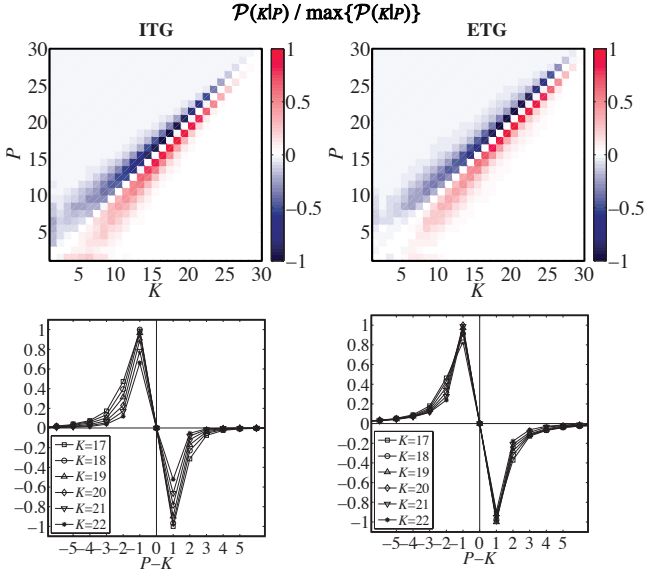}
\end{center}
\caption{(Color online) The shell-to-shell energy transfers for ITG and ETG driven turbulence. Below we plot the same normalised transfers, for given $\si$, as a function of $\sii-\si$.}
\label{fig_s2s_tot}
\end{figure}

\begin{figure*}[t]
\begin{center}
\includegraphics[width = 0.8\textwidth]{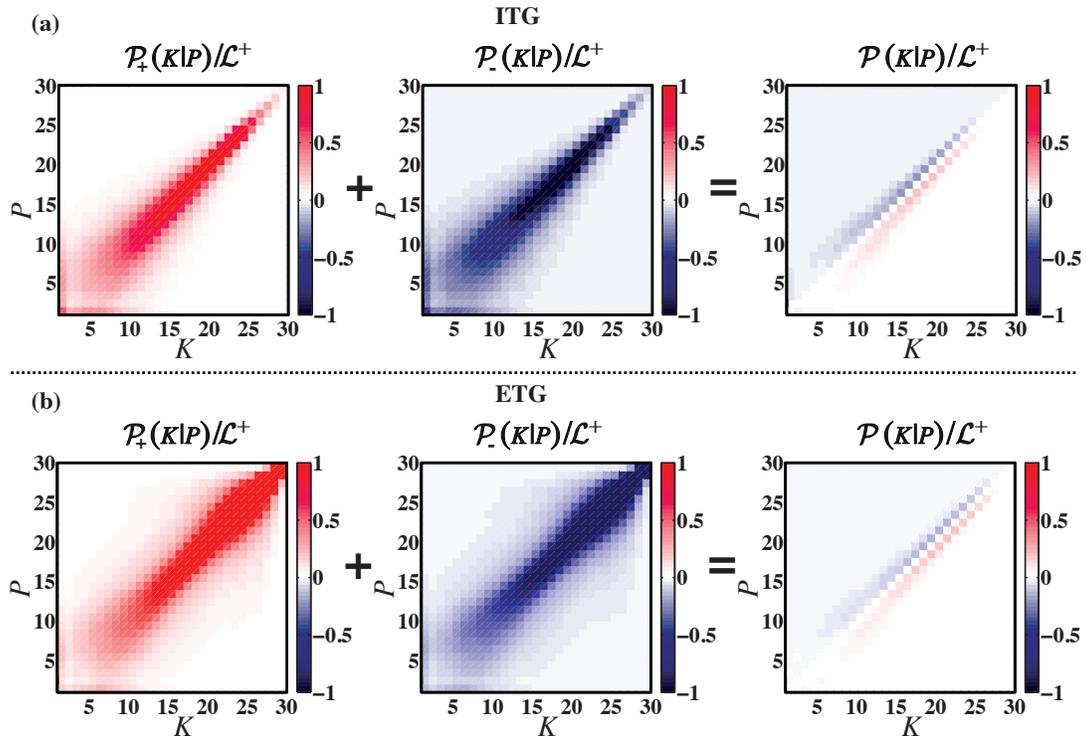}
\end{center}
\caption{(Color online) The shell-to-shell energy transfers for (a) ITG and (b) ETG driven turbulence. For each case, the forward and backward contributions are listed. The graphic representation shows that their sum recovers the net shell to shell transfer.}
\label{fig_s2s}
\end{figure*}
 
The scale-to-scale provides a diagnostic to visualize the energy cascade. Since the shell boundaries are taken as a power law, the normalized results to the maximal shell transfer provides us with information regarding the direction and locality of the energy cascade. We designate a transfer to be direct if it is positive for $\si>\sii$ and we call it local if $|\sii-\si|\sim5$ (due to our choice of $\lambda=2^{1/5}$). From FIG.~\ref{fig_s2s_tot}, we do observe that the scale-to-scale transfer pattern corresponds indeed to a direct and local energy cascade for ITG and ETG turbulence. Since $\mathcal P(\si|\sii)$ is systematically positive (lower-diagonal) for the energy received from larger scales $\si>\sii$, we can say that we observe a direct energy cascade.

It is important to differentiate between the locality of the energy cascade, one structure giving energy to a similar size structure, and the locality of interactions captured by the locality functions, where the mediator of the energetic interaction is also considered (to be presented in the next section). We stress that only the relative amplitude of the scale-to-scale transfers matters in comparing the intensity of the coupling between two scales.

\subsection{Forward and backward transfers}

The phenomenological interpretation of turbulence, as the direct cascade of energy from large to small scales, allows for the existence of backscatter. The backscatter represents transfer of information from small to large scales, in spite of an overall direct cascade. These transfers are believed to be important in the self-organization process of turbulence and are particular important in sub-grid scale modeling of turbulence.

Splitting the net transfers into a forward and backward component is arbitrary to a certain degree. Indeed, only at the mode-to-mode transfer level can we say with certainty that a transfer is directional (e.g., forward if $S(\kk|\pp|\qq)>0$ for $k<p$). At the shell level, as we are looking at effective transfers composed of many individual interaction, this becomes difficult. However, we use a similar approach as in Ref.~[\onlinecite{Debliquy:2005p203}] and propose a possible working definition. Considering the operators,
\begin{eqnarray}
\!\!\!\!\!\![X]_+=\left\{ \begin{array}{lcl}
X , &  \mbox{if}\ X> 0 \\
0 , &  \mbox{if}\ X\le0
\end{array}  \right.,\
[X]_-=\left\{ \begin{array}{lcl}
0 , &  \mbox{if}\ X> 0 \\
X , &  \mbox{if}\ X\le0
\end{array}  \right.,  
\end{eqnarray}
that select only the positive or negative contributions, we define the forward and backward triple-scale transfers as
\begin{align}
{\mathcal S}_+&(\si|\sii|\siii) \!=\frac{1}{2}\! \bigg{<}\! \bigg{[} \frac{T_0}{2F_{0}}\!\ \! \big{[} q_x p_y\!  - \! q_y  p_x \big{]} \phi^\siii\!(\qq) h^\sii\!(\pp)\bigg{]}_-  h^\si\!(\kk)  \nonumber \\
& - \bigg{[} \frac{T_0}{2F_{0}}\!\ \! \big{[} q_x p_y\!  - \! q_y  p_x \big{]} \phi^\siii\!(\qq) h^\si\!(\pp)\bigg{]}_+  h^\sii\!(\kk) \!\! \bigg{>}_\Lambda,
\label{s2s_forw}
\end{align} 
\begin{align}
{\mathcal S}_-&(\si|\sii|\siii) \!=\frac{1}{2}\! \bigg{<}\! \bigg{[} \frac{T_0}{2F_{0}}\!\ \! \big{[} q_x p_y\!  - \! q_y  p_x \big{]} \phi^\siii\!(\qq) h^\sii\!(\pp)\bigg{]}_+  h^\si\!(\kk)  \nonumber \\
& - \bigg{[} \frac{T_0}{2F_{0}}\!\ \! \big{[} q_x p_y\!  - \! q_y  p_x \big{]} \phi^\siii\!(\qq) h^\si\!(\pp)\bigg{]}_-  h^\sii\!(\kk) \!\! \bigg{>}_\Lambda.
\label{s2s_back}
\end{align}

The forward and backward decomposition is most useful when analyzing the scale-to-scale (shell-to-shell) transfers. The forward and backward scale-to-scale transfers are obtained from the triple-scale transfers, by summing over contribution made by all possible mediator scales $\siii$,
\begin{align}
\mathcal P_+(\si|\sii)=\sum_{\siii} \mathcal S^+(\si|\sii|\siii) \;, \nonumber\\
\mathcal P_-(\si|\sii)=\sum_{\siii} \mathcal S^-(\si|\sii|\siii) \;.
\label{def_s2s}
\end{align}
This definition allows for the property $\mathcal P_+(\si|\sii)=-\mathcal P_-(\sii|\si)$, which just tells that the contribution between two scales is equal and opposite. The net transfer value is recovered by summing the two contributions, $\mathcal P(\si|\sii)=\mathcal P_+(\si|\sii)+\mathcal P_-(\si|\sii)$.

In FIG.~\ref{fig_s2s} we plot the scale-to-scale transfers for the ITG and ETG cases. Since the forward and backward contributions are much larger than the net transfers, we use a different normalization, dividing everything by $\mathcal L^+$. This quantity represents the sum of the positive part of the linear contribution $\mathcal G + \mathcal L + \mathcal D$, 
\begin{align}
\mathcal L^+&= \int\limits\Big{[}\mathcal G(k) + \mathcal L (k)+ \mathcal D(k)\Big{]}_{+} \!\! \mbox{d}k \nonumber \\ 
&\approx \sum \Big{[}\mathcal G(\si) + \mathcal L (\si)+ \mathcal D(\si)\Big{]}_{+} \;,
\end{align}
here, for each case, the sum over the first 10 shells. 

From the start we notice that the net transfer results as the cancelation of two large contributions, significantly larger than that of the net transfer. Indeed, the anti-symmetry $\mathcal P_+(\si|\sii)=-\mathcal P_-(\sii|\si)$ recovers the $\mathcal P(\si|\sii)=-\mathcal P(\sii|\si)$ anti-symmetry property for the net transfer. This is important, since while the sign defined components would be easier to model, the models need to account properly for the difference between the two channels. Looking at the forward and backward components we still observe a local energy transfers, which occurs primarily between neighboring scales. Indeed, the transfers go to zero for $|\sii-\si|>5$. Since the geometric progression was taken with $\lambda=2^{1/5}$, the energy transfers can be seen as being local between octave-dyadic scales. This is true for the forward and backward decompositions as it is for the net transfer. 

Comparing the ITG and ETG cases, we do notice strong transfers at small scales for ETG turbulence, consistent with the linear transfer picture. This is important as a strong small scale cascade requires additional resolution to be properly solved. This ETG behavior was hinted at in other studies\cite{Nevins:2006p1454}, where resolutions considered sufficient for the ITG case led to a large difference in global value diagnostics (heat flux; not to be confused with scale-flux) for ETG turbulence.

\section{Scale fluxes perspective} \label{secV}

For turbulence, the flux of energy through a scale represents the most robust quantity related to the redistribution of energy. It is the quantity that stands at the basis of theoretical scaling arguments and phenomenological interpretations. Regardless of the definition, being based on the mode-to-mode-transfer or triad-transfers functions, the flux is consistent with the system dynamical symmetries and contains none of the uncertainties associated with the transfer functions.

\subsection{The free energy scale flux}

Formally, from the triad-transfers (Eq.~\ref{eq_triad}), we can define the flux of energy through a scale $k_\kc$ as
\begin{align}
\Pi(k_\kc) &=\! \bigg{<} \int\limits_{k\ge k_\kc}\!\!\!  \mbox{d}\kk \! \iint \! \mbox{d}\pp\, \mbox{d}\qq\ \mathcal T(\kk|\pp,\qq)\! \bigg{>}_\Theta \ .
\label{fluxdef}
\end{align}
From the wave-mode perspective, the physical interpretation is straightforward, as the energy received by modes $\kk$, located in a delimited zone of the wave-space, from the interaction with all other possible $\pp$ and $\qq$ modes. The wave-space is delimited by the surface $k_\kc$, designated here as a cut-off surface, through the condition $|\kk|\ge k_\kc$. In a slab geometry, the condition $|\kk|= k_\kc$ would trace a cylindric surface of radius $k_\kc$ in the $\{k_x,k_y, z\}$ space (the fluxes being through nested cylindrical surfaces). Due to the conservation of nonlinear interactions (Eq.~\ref{triad_conservation}), only interactions that cross the $k_\kc$ surface make a non-zero contribution to the flux. This allows the flux to be seen as the energy transferred through a wave-surface, rather than energy transferred between wave-modes. 

Numerically, we have access to the triple-scale transfer $\mathcal S(\si|\sii|\siii)$ and thus, we can compute with ease the scale flux through the scales boundaries ($k_\kc=k_\si$). In term of the triple-scale transfer information, the scale flux reads as
\begin{align}
\Pi(k_\kc) &=\sum_{\si=c+1}^{N} \sum_{\sii=1}^{N} \sum_{\siii=1}^{N} \mathcal S(\si|\sii|\siii) \nonumber \\
&=\sum_{\si=c+1}^{N} \sum_{\sii=1}^{N}  \mathcal P(\si|\sii) \nonumber \\
&=\sum_{\si=c+1}^{N} \mathcal T(\si) \label{fluxdeff}\ , 
\end{align}
where the last two identities relate the scale flux to the scale-to-scale and transfer-spectra quantities, respectively. These relations, obvious from the definitions of scale-to-scale (Eq.~\ref{s2s_def}) and transfer-spectra (Eq.~\ref{lt_def}), are meant to emphasize the level from which the scale flux information can be recovered. Integrating over the total energy received by a scale is sufficient, as all possible triad contributions are automatically taken into account. 
 
\begin{figure}[tb]
\begin{center}
\includegraphics[width = 0.48\textwidth]{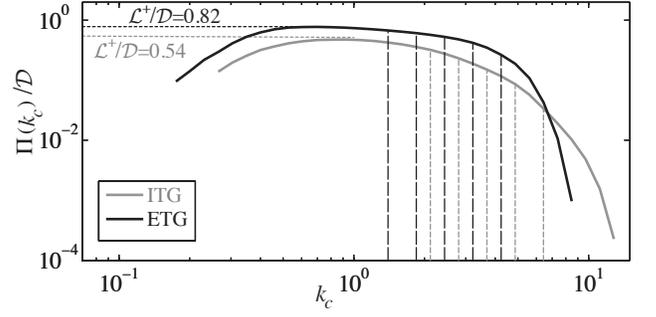}
\end{center}
\caption{The contributions to the free-energy flux across $k_\kc$ for ITG and ETG cases. For each case separately, $k_\kc$ is normalised to the respective gyroradius. The vertical dashed lines represent the upper boundary of shells $c=\{16, 18, 20, 22, 24\}$ for the respective case.}
\label{fig_flux_ITG_ETG}
\end{figure}
 
For the two cases of GK turbulence studied here, ITG and ETG, we plot in FIG.~\ref{fig_flux_ITG_ETG} the flux of energy through a scale. While in both cases the flux value is increasing slowly in the scale range dominated by the injection of energy (the first $10$ shells for each respective case), the ETG flux exhibits a tendency to level in value over a certain range, before quickly decreasing to zero. In comparison, the ITG flux presents a more gradual transition from the injection dominated build up of the flux to the dissipation dominated decrease in value. The presence of a constant flux over a certain range of scales, i.e., an inertial range, represents the call sign of a classical turbulence behavior. Even if for sufficiently separated injection and dissipation ranges one would expect the ITG flux to develop a similar plateau (the authors were unable to verify this assumption with the resolutions and computational resources available), we can clearly say that ETG turbulence behaves in a more classical way than ITG. 

Another sign that supports this assessment consists in the flux saturation value. In classical turbulence, the energy injection rate equals the energy dissipation value and that of the scale flux. While in GK turbulence, at the global level, the total energy injected still balances the total energy dissipated for a steady state, the scale flux is not saturated by this value. As dissipation can act strongly at the same range as the injection of energy, only a ratio of the energy injected gets transferred. This value is designated here as $\mathcal L^+$ and represents the sum of the positive part of the linear contribution $\mathcal G + \mathcal L + \mathcal D$, here the sum over the first 10 shells. A larger $\mathcal L^+/\mathcal D$ ratio, tending to one, denotes that a larger amount of energy injected in the system is passed down to the turbulent cascade and thus, turbulence behaves closer to the classical picture.

\subsection{Flux contributions}

\begin{figure}[b]
\begin{center}
\includegraphics[width = 0.38\textwidth]{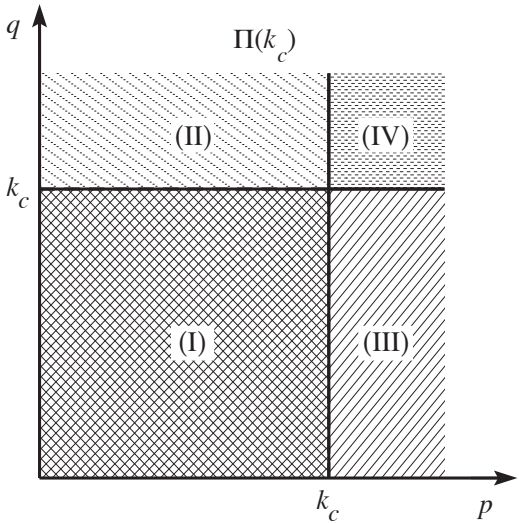}
\end{center}
\caption{The contributions to the free energy flux across $k_\kc$, from $p$ and $q$.}
\label{fig_flux_def}
\end{figure}

To better understand the triad transfers contributing to the energy flux in Eq.~(\ref{fluxdef}), we decompose the last two integrals taken over $\pp$ and $\qq$ in respect to $k_\kc$ and we label the four terms to ease their identifications,
\begin{align}
\label{flx2}
\Pi(k_\kc)&=\bigg{<}\int\limits_{k\ge k_\kc} \!\!\!  \mbox{d}\kk \Bigg{[} \underbrace{\ \int\limits_{p< k_\kc} \!\!\!  \mbox{d}\pp  \!\! \int\limits_{q< k_\kc} \!\!\!  \mbox{d}\qq}_{I}\ \ +\!\!\!\! \underbrace{\ \int\limits_{p< k_\kc} \!\!\!  \mbox{d}\pp \!\! \int\limits_{q\ge k_\kc} \!\!\!  \mbox{d}\qq}_{II}\ \ +\nonumber \\
&+\!\!\!\! \underbrace{\ \int\limits_{p\ge k_\kc} \!\!\!  \mbox{d}\pp  \!\! \int\limits_{q< k_\kc} \!\!\!  \mbox{d}\qq}_{III}\ \ +\!\!\!\! \underbrace{\ \int\limits_{p\ge k_\kc} \!\!\!  \mbox{d}\pp \!\! \int\limits_{q\ge k_\kc} \!\!\!  \mbox{d}\qq}_{IV} \Bigg{]} \mathcal T(\kk|\pp,\qq)\! \bigg{>}_\Theta\;.
\end{align}
The contributions of the terms, in respect to $p$ and $q$ scales, are represented schematically in FIG.~\ref{fig_flux_def}. The first term $(I)$ contains the contribution of triads which have both legs across the surface. For the second term ($II$), only $p$ is across the cutoff surface, while for the third term ($III$) only the $q$ leg of the triad penetrates the surface. Since these two terms are equal in contribution, as the triad transfer $\mathcal T(\kk|\pp,\qq)$ is symmetric in $\pp$ and $\qq$, we will look at their sum. As mentioned, the last term ($IV$) is always zero due to the conservation of interactions. We mention that the contributions made by the term $I$ and the sum $II+III$ are the same when performing a similar decomposition on the triple-scale transfer $\mathcal S(\si|\sii|\siii)$, even if the resulting individual terms $II$ and $III$ are not. This is due to the $\qq$ and $\pp$ symmetry breaking for $\mathcal S(\si|\sii|\siii)$ [see Eq.~(\ref{eq_triad_brk}) and Eq.~(\ref{eq_tripleshell})], resulting from the filtering procedure, a symmetry that is restored by the sum of the two terms ($II+III$). 

\begin{figure}[tb]
\begin{center}
\includegraphics[width = 0.43\textwidth]{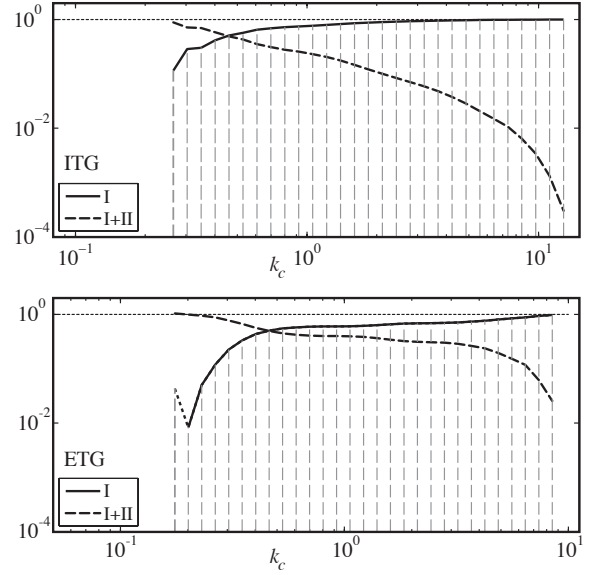}
\end{center}
\caption{The free energy flux nonzero contributions ($I$ and $II+III$), normalized to the total flux, for the ITG and ETG cases. For ETG, the fist shell boundary contribution of term $I$ is negative. In both cases, the vertical dash lines represent the respective shell boundaries.}
\label{fig_flux_split}
\end{figure}

In FIG.~\ref{fig_flux_split} we plot the nonzero flux contributions ($I$ and $II+III$) as a ratio of the total flux, for the ITG and ETG cases. It can be seen that for the ETG case the $II+III$ contributions decrease more slowly, showing a tendency to level out. A constant ratio between these two flux contributions is to be expected in an infinitely long self-similar dynamical range (i.e. the inertial range). More importantly, as we will see next, even if the first term ($I$) is responsible for the main contribution to the flux, or has a comparable contribution to the sum $II+III$, the locality of interaction of the two contributions is drastically different.

\subsection {Infrared (IR) locality functions}

In general, the idea of locality can be seen as the disparity between scales contributing to a nonlinear interaction~\cite{Eyink:2005p854}. For a given energy flux through a scale, the degree to which each scale contributes to the mentioned flux represents an assertion of locality. For the interaction to be local, the contribution of highly separated scales should be small and decrease fast with the increase in separation\cite{Teaca:2011p1362}. Measuring the locality through the perspective of the scale flux can be done using Kraichnan's infrared (IR) and ultraviolet (UV) locality functions\cite{Kraichnan:1959p497}.

Starting from the definition of the flux, the IR locality function is defined by taking a second probe surface ($k_\kp$) in such a way ($k_\kp \le k_\kc$) that it limits the selection of triads that contribute to the energy flux through $k_\kc$. Conceptually, the definition can be obtained from Eq.~(\ref{flx2}) (the nonzero terms), by replacing the integral limits inside the square bracket from $k_\kc$ to $k_\kp$ and it reads as 
\begin{align}
\label{IRloc}
\Pi_{\mbox{\scriptsize ir}}(k_\kp|k_\kc)= \bigg{<}
\int\limits_{k\ge k_\kc} \!\!\!  \mbox{d}\kk \Bigg{[}& \ \int\limits_{p< k_\kp} \!\!\! \mbox{d}\pp \!\! \int\limits_{q< k_\kp} \!\!\!  \mbox{d}\qq \ +\nonumber \\
&2\!\!\! \int\limits_{p< k_\kp} \!\!\!  \mbox{d}\pp \!\! \int\limits_{q\ge k_\kp} \!\!\!  \mbox{d}\qq\Bigg{]} \mathcal T(\kk|\pp,\qq)\! \bigg{>}_\Theta \;.
\end{align}
It measures the contribution to the flux through $k_\kc$ from triads of modes with at least one scale larger than that of the probe $k_\kp$. In the limit $k_\kp \rightarrow k_\kc$, the locality functions recover the value of the flux across the cut-off $k_\kc$. This allows for the normalization $\Pi_{\mbox{\scriptsize ir}}(k_\kp|k_\kc)/\Pi(k_\kc)$ to be one for $k_\kp=k_\kc$. Moreover, since for a steady turbulent state the fluxes are directional (same sign across a range of scales), for $k_\kp<k_\kc$ the normalized value of the locality functions will decrease in value as a function of $k_\kp/k_\kc$. Thus, we can measure the rate at which the normalized value of the locality functions decreases as a function of separation between the two scales (identified by $k_\kp$ and $k_\kc$). Larger rates than otherwise imply that contributions made by ever separated scales are smaller and thus local interactions dominate. 

\begin{figure}[tb]
\begin{center}
\includegraphics[width = 0.40\textwidth]{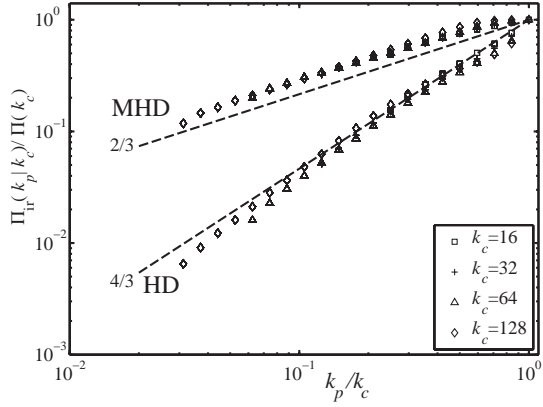}
\end{center}
\caption{IR locality functions obtained for hydrodynamical (HD) and magneto-hydrodynamical (MHD) turbulence data~\cite{Domaradzki:2009p557, Domaradzki:2010p1012, Teaca:2011p1362}. In both cases the same $512^3$ resolution, simulation parameters and forcing mechanism has been used.}
\label{fig_IR_HD_MHD}
\end{figure}

To better understand these results, we judge them from the premiss of classical turbulence, for an infinitely long inertial range. The collapse of the normalized locality functions curves, for different $k_\kc$ values, implies a self-similarity behavior of the dynamics (although not necessarily given by a simple scaling law). Furthermore, collapsing on the same $(k_\kp/k_\kc)^{\alpha}$ curve denotes a universal nonlinear interaction regime consistent with a simple scaling law for turbulence. The exponent $\alpha$ is known as asymptotic locality exponent, as higher degrees of turbulence will not generate larger nonlocal interactions. This simple picture is found for hydrodynamical (HD) and magneto-hydrodynamical (MHD) turbulence. The theoretical locality exponents $4/3$ and $2/3$, respectively, have been confirmed numerically\cite{Zhou:1993p949, Domaradzki:2009p557, Eyink:2009p808, Aluie:2009p809}, the latter case referring to the total energy flux contribution\cite{Teaca:2011p1362}. For reference, we plot in FIG.~\ref{fig_IR_HD_MHD} a composite figure of the IR locality functions for HD and MHD turbulence obtained from data used by the authors in their previous works~\cite{Domaradzki:2009p557, Domaradzki:2010p1012, Teaca:2011p1362}.
  
\begin{figure}[b]
\begin{center}
\includegraphics[width = 0.38\textwidth]{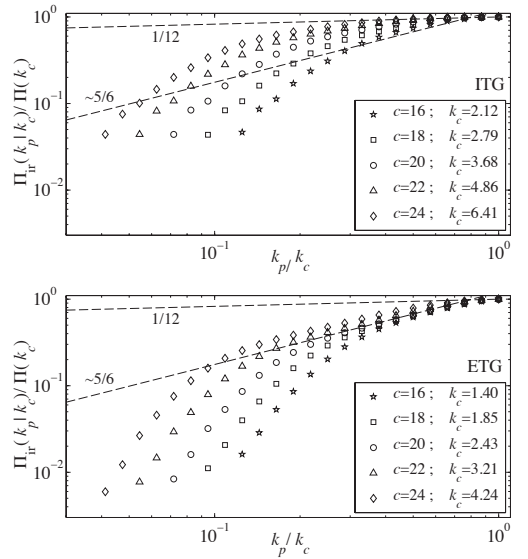}
\end{center}
\caption{IR locality functions for ITG and ETG turbulence, normalised to the value of the total flux in each case. The cut-off values $k_\kc$ correspond to the ones identified in FIG.~\ref{fig_flux_def} by the vertical dashed lines.}
\label{fig_IR}
\end{figure}

In FIG.~\ref{fig_IR} we plot the locality functions for different cut-off values $k_\kc$. While the values of $k_\kc$ chosen for ITG and ETG cases differ, their location in respect to the peak of the flux is similar (same shells values $c$ are desired for a comparison of turbulence behavior). Compared to the simpler classical turbulence examples mentioned above, GK turbulence struggles to recover a similar picture. First we need to consider the smaller resolutions available that would limit the appearance of a large "inertial range" behavior. Our choice in cut-off limits ($c$ values selected for the IR functions displayed in FIG.~\ref{fig_flux_ITG_ETG}) are meant to span the plateau range of the flux, while avoiding the strongly damped scales. Second, as the gradient driven turbulence has a wide range dominated by the driving instability, we notice that the curves exhibit a similar fallout once this zone is reached, regardless of the cut-off. In this unclear situation, we look for signs of asymptotic locality, curves collapsing on the same slope. For the ETG case, we see a tendency to recover the theoretical 5/6 value~\cite{Teaca:2012p1415}. This is particularly encouraging as in this case, we do notice a tendency of the curves to collapse on each other (here, $c=16$, $c=18$ and $c=20$ in particular). By comparison, a wide range of values are found for ITG, ranging from the 5/6 to the 1/12 value. For ITG turbulence, we do not observe the collapse of the IR functions on any value, regardless of the $k_\kc$ selection.

\begin{figure}[b]
\begin{center}
\includegraphics[width = 0.48\textwidth]{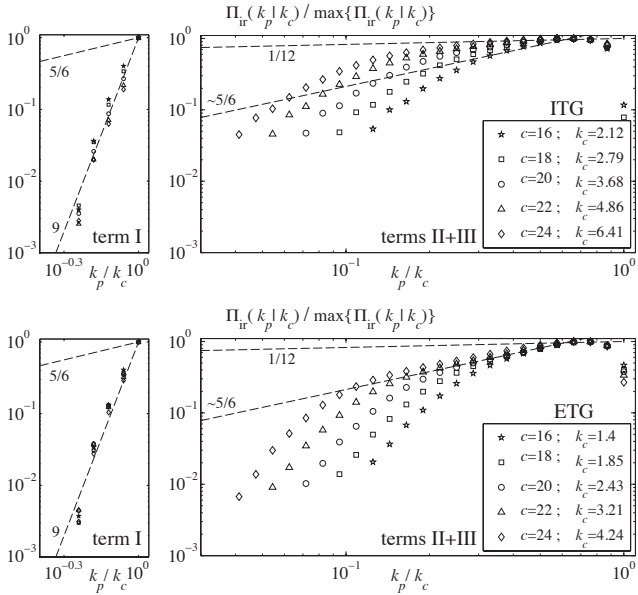}
\end{center}
\caption{Contribution of the $I$ and $II+III$ terms to the IR locality for ITG and ETG turbulence. Each function is normalised to its maximal value, to collapse the curves onto each other.}
\label{fig_IR_III}
\end{figure}

To aid our understanding, we look (FIG.~\ref{fig_IR_III}) at the locality functions of the $I$ and $II+III$ contributions to the flux, corresponding to the fist and second term in Eq.~(\ref{IRloc}), respectively. Each function is normalized to its maximal value, to collapse the curves onto each other, as we are interested in observing the slope behavior. For term $I$, in both cases, an ultra local behavior is found (a line with an exponent of $9$ is drawn for reference), the locality functions exponent dropping to zero over a $5$ shell interval. It should be pointed out that this strong local nature of the dominant flux contribution needs to be exploited in future modeling attempts. Thus the main influence to the locality exponent is made by the $II+III$ flux contribution. This is not surprising, as triads with only one leg through the cut-off surface are geometrically less constrained and allow for the most non-local interactions (seen phenomenologically as the shearing of small scale structures by large scale fluctuations, as discussed in detail in Ref.~[\onlinecite{Howes:2011p1459}]). The ETG case tendency to collapse on the 5/6 slope is again evident as is the undetermined exponent for ITG turbulence. The cause of these different behaviors is explored next.

\subsection {Zonal flow contribution on the locality functions}

\begin{figure}[h]
\begin{center}
\includegraphics[width = 0.40\textwidth]{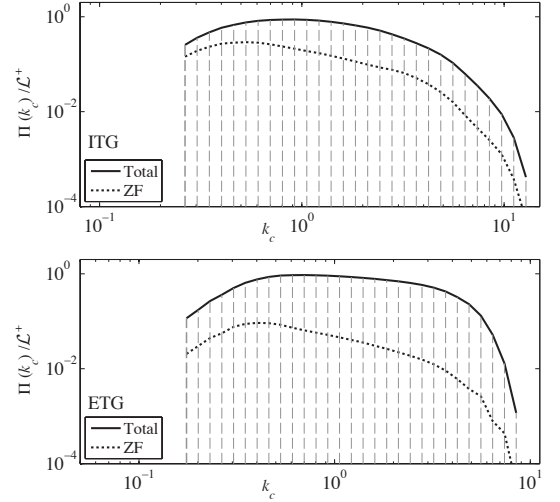}
\end{center}
\caption{Zonal flow modes scale flux contributions in respect to the total flux for ITG and ETG turbulence.}
\label{fig_ZF_flux}
\end{figure}

While ETG and ITG behaves differently, we still need to understand why. Using $\mathcal S^{ZF}(\si|\sii|\siii)$ as the building block in Eq.~(\ref{fluxdeff}), we construct the flux through a scale $\Pi^{ZF}(k_\kc)$ and the coresponding infrared locality function due to the zonal-flow modes $\kk^{ZF}$ (here $k_y=0$ modes). In general, the total energy flux can be decomposed into a series of fluxes that only contain interactions that possess certain symmetry constraints.  
\begin{align}
\Pi(k_\kc)=\Pi^{ZF}(k_\kc)+\Pi^{\mbox{remainder}}(k_\kc).
\end{align}
In FIG.~\ref{fig_ZF_flux}, we plot the flux contributions. We observe that the zonal-flow contribution is smaller for the ETG case. This is important since regardless of their non-locality behavior, the degree to which such contributions affect the overall picture depends on their amplitude.

\begin{figure}[t]
\begin{center}
\includegraphics[width = 0.40\textwidth]{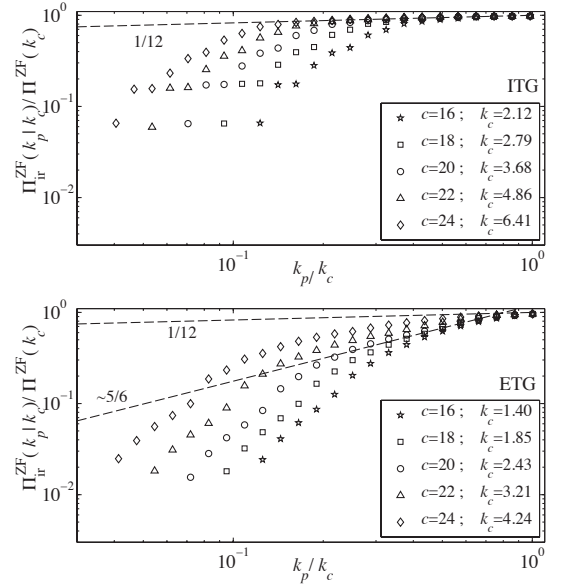}
\end{center}
\caption{IR locality functions for ITG and ETG zonal-flow modes contribution, normalized to their respective scale flux. The cut-off values $k_\kc$ correspond to the ones identified in FIG.~\ref{fig_flux_def} by the vertical dashed lines.}
\label{fig_ZF_IR}
\end{figure}

In FIG.~\ref{fig_ZF_IR} we plot the IR locality functions arising from the interactions involving the zonal-flow modes ($\kk^{ZF}$). We see a clear $1/12$ exponent for ITG, a very non-local contribution. This seems to indicate the zonal-flow modes contribution in the case of ITG (emphasized due to the electron adiabatic response) as the cause of the more pronounced non-local nature compared to ETG turbulence. 
 
\subsection {Ultraviolet (UV) locality functions}

Since $k_\kc$ is fixed at a given value, the IR locality functions let us know if the energy transferred by the flux $\Pi(k_\kc)$ comes primarily from nearby larger scales (a local behavior) or from scales $k_\kp$ located farther apart (a more non-local behavior). However, these functions do not possess information on where the energy is deposited across the flux cut-off $k_\kc$. This information is given by the ultraviolet (UV) locality functions. The UV locality functions are obtained in an analog way to the IR ones, starting from the definition of the flux,
\begin{align}
\Pi(k_\kc) &=\! \bigg{<} \!\!  -\!\!\!\! \int\limits_{k\le k_\kc}\!\!\! \mbox{d}\kk \! \iint \! \mbox{d}\pp\, \mbox{d}\qq\  \mathcal T(\kk|\pp,\qq)\! \bigg{>}_\Theta \ ,
\label{fluxdef2}
\end{align}
obtained from Eq.~(\ref{fluxdef}) by considering the conservation of interactions in a triad (changing the integral limit to $k\le k_\kc$ introduces the minus sign). Decomposing the last two integrals taken over $\pp$ and $\qq$ in respect to $k_\kc$ and taking $k_\kp \ge k_\kc$ gives the ultraviolet (UV) locality functions definition, 
\begin{align}
\label{UVloc}
\Pi_{\mbox{\scriptsize uv}}(k_\kp|k_\kc)= \bigg{<}
\int\limits_{k< k_\kc} \!\!\!  \mbox{d}\kk \Bigg{[} &\ \int\limits_{p\ge k_\kp} \!\!\!  \mbox{d}\pp \!\! \int\limits_{q\ge k_\kp} \!\!\!  \mbox{d}\qq \ +\nonumber \\
&2\!\!\! \int\limits_{p< k_\kp} \!\!\!  \mbox{d}\pp \!\! \int\limits_{q\ge k_\kp} \!\!\!  \mbox{d}\qq\Bigg{]} \mathcal T(\kk|\pp,\qq)\! \bigg{>}_\Theta \;.
\end{align}
%

\begin{figure}[tb]
\begin{center}
\includegraphics[width = 0.43\textwidth]{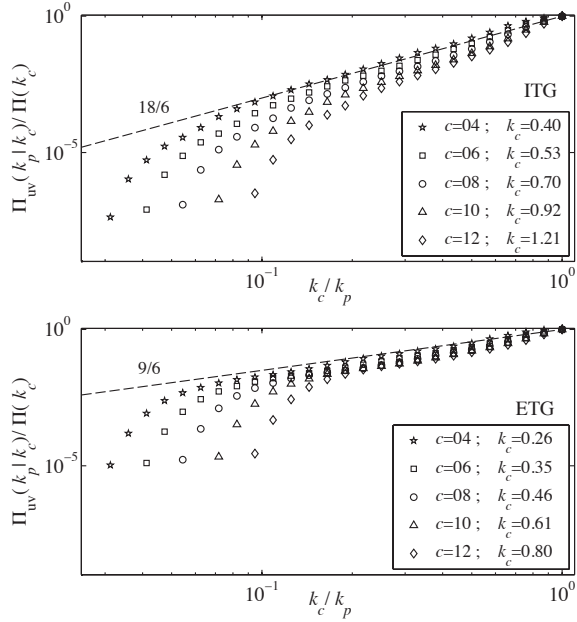}
\end{center}
\caption{UV locality functions for ITG and ETG turbulence, normalised to the value of the total flux in each case.}
\label{fig_locUV}
\end{figure}

The functions measure the contribution to the flux through $k_\kc$ from triads of modes with at least one wavenumber greater than $k_\kp$, therefore providing information regarding the locality makeup of a scale $k_\kc$ in relation with smaller and smaller scales ($k_\kc/k_\kp \rightarrow 0$). Thus, ignoring where the energy is coming from through $k_\kc$, we can measure where the energy is primarily deposited by looking at the $\Pi_{\mbox{\scriptsize ir}}(k_\kp|k_\kc)/\Pi(k_\kc)$ ratio in respect to $k_\kc/k_\kp$.

Since determining asymptotic locality exponents requires us to span the inertial range, while being out of the driving range influence, the UV locality exponent is much harder to determine numerically. This is a known problem even for the simpler HD and MHD turbulent systems. In FIG.~\ref{fig_locUV} we present the normalised UV locality functions for ITG and ETG turbulence. In both cases we see a departure from the theoretical $5/6$ exponent estimate. For ITG turbulence, we observe an upper value of $18/6$. In the ETG case, while we do observe a $9/6$ value, we also observe a tendency for the functions to collapse on the same slope. Although both values are interpreted by us as a local UV makeup of interactions, we do see that ETG turbulence is less local than ITG. This is consistent with the idea of a stronger cascade, requiring a larger range of scales to be fully accounted for.

\section {Conclusions and discussions}

In this work, the energy exchanges between perpendicular spatial scales were analysed for ITG and ETG driven gyrokinetic turbulence. From the start, the tendency of ETG turbulence to experience higher value small-scale transfers is evident from the transfer spectra. This effect is further confirmed from the perspective of the scale-to-scale interactions, including the forward and backward decompositions. While in both cases, we see a direct, local transfer of energy, the increase in small-scale scale transfers for ETG turbulence denotes a stronger cascade. A quantifiable diagnostic to this effect is given at the flux level, where $82\%$ of the energy injected in the ETG system is cascaded down compared to only  $54\%$ for ITG.

The strength of cascade has an influence on the turbulence properties. ETG has a stronger classical turbulent behavior, recovering the theoretical IR locality exponent. In the ITG case, no clear exponent is found. This is due in part to the strongly non-local zonal flow component, but also to its reduced cascade that does not redistribute a large enough value of energy to reach a set of scales unaffected by this large scale effect. Compared to ETG, the stronger non-local character of ITG implies that the cascade (seen as being local from the scale-to-scale exchanges) depends more on the energy exchanges mediated by large scales than on those mediated by scales comparable in size.

From a modeling perspective, these two different behaviors have non-intuitive implications. As with all turbulent systems, it is desired to model the small-scale effects while numerically computing only the largest scales (eddies) in the system. Such an approach is taken by the Large Eddy Simulations (LES) sub-grid scale modeling technique. This method requires the existence of universal small scales that, more importantly, ensure a universal cascading mechanism in the form of an inertial range. In the inertial range (towards its beginning), a cut-off is taken and scales smaller than that are removed while their effects on larger scales are accounted for by a model. For this model to be universal, the removed small scales should have a small dependence on the large scales. This information is given by the IR locality functions. Thus, the more local ETG turbulence behavior is preferred. Naturally, the impact made by a model on the resolved large scales needs to be consistent with the impact made by the small scales on the large ones. However, this information is given by the UV locality functions. From this perspective, while generating an ITG model is more problematic, its impact on the system is less important than for ETG (UV locality being higher for ITG than ETG). This effect was seen in a separate study related to the application of LES to gyrokinetics\cite{BanonNavarro:2014p1535}.

Non-intuitively, the same complications that make ITG harder to understand and model from the perspective of classical turbulence, also makes the  system more robust to scale truncations. Conversely, ETG turbulence is more dependent on numerical resolution than ITG, experiencing a stronger cascade that needs to be fully resolved, while at the same time being a perfect candidate for the implementation of LES methods and the use of classical turbulence scaling arguments. 

At the end we mention that for a conclusive understanding of the fundamental properties of GK turbulence, the impact of the resolution and of the geometry needs to be understood from the perspective of the fundamental energy transfer in the five-dimensional space. Due to the computational resources required and the hurdles of a full spectral implementation, such progress is limited, but should not be abandoned by the community.

\acknowledgments
The authors would like to thank T. G\"orler, D. Told and A. Bustos for fruitful discussions. BT would like to thank Stephan Brunner and Laurent Villard, both from EPFL-CRPP Switzerland, for discussions regarding gyrokinetic turbulence energetics, while he was stationed there. We gratefully acknowledge that the results in this paper have been achieved with the assistance of high performance computing of the HELIOS system  hosted at the International Fusion Energy Research Centre (IFERC) in Japan. The research leading to these results received  funding from the European Research Council under the European Unions Sevenths Framework Programme (FP7/2007-2013) / ERC Grant Agreement No.~277870.


\begin{thebibliography}{35}%
\makeatletter
\providecommand \@ifxundefined [1]{%
 \@ifx{#1\undefined}
}%
\providecommand \@ifnum [1]{%
 \ifnum #1\expandafter \@firstoftwo
 \else \expandafter \@secondoftwo
 \fi
}%
\providecommand \@ifx [1]{%
 \ifx #1\expandafter \@firstoftwo
 \else \expandafter \@secondoftwo
 \fi
}%
\providecommand \natexlab [1]{#1}%
\providecommand \enquote  [1]{``#1''}%
\providecommand \bibnamefont  [1]{#1}%
\providecommand \bibfnamefont [1]{#1}%
\providecommand \citenamefont [1]{#1}%
\providecommand \href@noop [0]{\@secondoftwo}%
\providecommand \href [0]{\begingroup \@sanitize@url \@href}%
\providecommand \@href[1]{\@@startlink{#1}\@@href}%
\providecommand \@@href[1]{\endgroup#1\@@endlink}%
\providecommand \@sanitize@url [0]{\catcode `\\12\catcode `\$12\catcode
  `\&12\catcode `\#12\catcode `\^12\catcode `\_12\catcode `\%12\relax}%
\providecommand \@@startlink[1]{}%
\providecommand \@@endlink[0]{}%
\providecommand \url  [0]{\begingroup\@sanitize@url \@url }%
\providecommand \@url [1]{\endgroup\@href {#1}{\urlprefix }}%
\providecommand \urlprefix  [0]{URL }%
\providecommand \Eprint [0]{\href }%
\providecommand \doibase [0]{http://dx.doi.org/}%
\providecommand \selectlanguage [0]{\@gobble}%
\providecommand \bibinfo  [0]{\@secondoftwo}%
\providecommand \bibfield  [0]{\@secondoftwo}%
\providecommand \translation [1]{[#1]}%
\providecommand \BibitemOpen [0]{}%
\providecommand \bibitemStop [0]{}%
\providecommand \bibitemNoStop [0]{.\EOS\space}%
\providecommand \EOS [0]{\spacefactor3000\relax}%
\providecommand \BibitemShut  [1]{\csname bibitem#1\endcsname}%
\let\auto@bib@innerbib\@empty
\bibitem [{\citenamefont {Krommes}(2012)}]{Krommes:2012p1373}%
  \BibitemOpen
  \bibfield  {author} {\bibinfo {author} {\bibfnamefont {J.~A.}\ \bibnamefont
  {Krommes}},\ }\href {\doibase 10.1146/annurev-fluid-120710-101223} {\bibfield
   {journal} {\bibinfo  {journal} {Annual Review of Fluid Mechanics}\ }\textbf
  {\bibinfo {volume} {44}},\ \bibinfo {pages} {175} (\bibinfo {year}
  {2012})}\BibitemShut {NoStop}%
\bibitem [{\citenamefont {Brizard}\ and\ \citenamefont
  {Hahm}(2007)}]{Brizard:2007p11}%
  \BibitemOpen
  \bibfield  {author} {\bibinfo {author} {\bibfnamefont {A.~J.}\ \bibnamefont
  {Brizard}}\ and\ \bibinfo {author} {\bibfnamefont {T.~S.}\ \bibnamefont
  {Hahm}},\ }\href {\doibase 10.1103/RevModPhys.79.421} {\bibfield  {journal}
  {\bibinfo  {journal} {Rev. Mod. Phys.}\ }\textbf {\bibinfo {volume} {79}},\
  \bibinfo {pages} {421} (\bibinfo {year} {2007})}\BibitemShut {NoStop}%
\bibitem [{\citenamefont {Hammett}, \citenamefont {Dorland},\ and\
  \citenamefont {Perkins}(1992)}]{Hammett:1992p1538}%
  \BibitemOpen
  \bibfield  {author} {\bibinfo {author} {\bibfnamefont {G.~W.}\ \bibnamefont
  {Hammett}}, \bibinfo {author} {\bibfnamefont {W.}~\bibnamefont {Dorland}}, \
  and\ \bibinfo {author} {\bibfnamefont {F.~W.}\ \bibnamefont {Perkins}},\
  }\href {\doibase 10.1063/1.860014} {\bibfield  {journal} {\bibinfo  {journal}
  {Physics of Fluids B: Plasma Physics}\ }\textbf {\bibinfo {volume} {4}},\
  \bibinfo {pages} {2052} (\bibinfo {year} {1992})}\BibitemShut {NoStop}%
\bibitem [{\citenamefont {Watanabe}\ and\ \citenamefont
  {Sugama}(2006)}]{Watanabe:2006p1444}%
  \BibitemOpen
  \bibfield  {author} {\bibinfo {author} {\bibfnamefont {T.-H.}\ \bibnamefont
  {Watanabe}}\ and\ \bibinfo {author} {\bibfnamefont {H.}~\bibnamefont
  {Sugama}},\ }\href {\doibase 10.1088/0029-5515/46/1/003} {\bibfield
  {journal} {\bibinfo  {journal} {Nucl. Fusion}\ }\textbf {\bibinfo {volume}
  {46}},\ \bibinfo {pages} {24} (\bibinfo {year} {2006})}\BibitemShut {NoStop}%
\bibitem [{\citenamefont {Schekochihin}\ \emph {et~al.}(2008)\citenamefont
  {Schekochihin}, \citenamefont {Cowley}, \citenamefont {Dorland},
  \citenamefont {Hammett}, \citenamefont {Howes}, \citenamefont {Plunk},
  \citenamefont {Quataert},\ and\ \citenamefont
  {Tatsuno}}]{Schekochihin:2008p1034}%
  \BibitemOpen
  \bibfield  {author} {\bibinfo {author} {\bibfnamefont {A.~A.}\ \bibnamefont
  {Schekochihin}}, \bibinfo {author} {\bibfnamefont {S.~C.}\ \bibnamefont
  {Cowley}}, \bibinfo {author} {\bibfnamefont {W.}~\bibnamefont {Dorland}},
  \bibinfo {author} {\bibfnamefont {G.~W.}\ \bibnamefont {Hammett}}, \bibinfo
  {author} {\bibfnamefont {G.~G.}\ \bibnamefont {Howes}}, \bibinfo {author}
  {\bibfnamefont {G.~G.}\ \bibnamefont {Plunk}}, \bibinfo {author}
  {\bibfnamefont {E.}~\bibnamefont {Quataert}}, \ and\ \bibinfo {author}
  {\bibfnamefont {T.}~\bibnamefont {Tatsuno}},\ }\href {\doibase
  10.1088/0741-3335/50/12/124024} {\bibfield  {journal} {\bibinfo  {journal}
  {Plasma Phys. Control. Fusion}\ }\textbf {\bibinfo {volume} {50}},\ \bibinfo
  {pages} {4024} (\bibinfo {year} {2008})}\BibitemShut {NoStop}%
\bibitem [{\citenamefont {Tatsuno}\ \emph {et~al.}(2009)\citenamefont
  {Tatsuno}, \citenamefont {Dorland}, \citenamefont {Schekochihin},
  \citenamefont {Plunk}, \citenamefont {Barnes}, \citenamefont {Cowley},\ and\
  \citenamefont {Howes}}]{Tatsuno:2009p1096}%
  \BibitemOpen
  \bibfield  {author} {\bibinfo {author} {\bibfnamefont {T.}~\bibnamefont
  {Tatsuno}}, \bibinfo {author} {\bibfnamefont {W.}~\bibnamefont {Dorland}},
  \bibinfo {author} {\bibfnamefont {A.~A.}\ \bibnamefont {Schekochihin}},
  \bibinfo {author} {\bibfnamefont {G.~G.}\ \bibnamefont {Plunk}}, \bibinfo
  {author} {\bibfnamefont {M.}~\bibnamefont {Barnes}}, \bibinfo {author}
  {\bibfnamefont {S.~C.}\ \bibnamefont {Cowley}}, \ and\ \bibinfo {author}
  {\bibfnamefont {G.~G.}\ \bibnamefont {Howes}},\ }\href {\doibase
  10.1103/PhysRevLett.103.015003} {\bibfield  {journal} {\bibinfo  {journal}
  {Phys. Rev. Lett.}\ }\textbf {\bibinfo {volume} {103}},\ \bibinfo {pages}
  {15003} (\bibinfo {year} {2009})}\BibitemShut {NoStop}%
\bibitem [{\citenamefont {Plunk}\ \emph {et~al.}(2010)\citenamefont {Plunk},
  \citenamefont {Cowley}, \citenamefont {Schekochihin},\ and\ \citenamefont
  {Tatsuno}}]{Plunk:2010p1360}%
  \BibitemOpen
  \bibfield  {author} {\bibinfo {author} {\bibfnamefont {G.~G.}\ \bibnamefont
  {Plunk}}, \bibinfo {author} {\bibfnamefont {S.~C.}\ \bibnamefont {Cowley}},
  \bibinfo {author} {\bibfnamefont {A.~A.}\ \bibnamefont {Schekochihin}}, \
  and\ \bibinfo {author} {\bibfnamefont {T.}~\bibnamefont {Tatsuno}},\ }\href
  {\doibase 10.1017/S002211201000371X} {\bibfield  {journal} {\bibinfo
  {journal} {J. Fluid Mech.}\ }\textbf {\bibinfo {volume} {664}},\ \bibinfo
  {pages} {407} (\bibinfo {year} {2010})}\BibitemShut {NoStop}%
\bibitem [{\citenamefont {Barnes}, \citenamefont {Parra},\ and\ \citenamefont
  {Schekochihin}(2011)}]{Barnes:2011p1372}%
  \BibitemOpen
  \bibfield  {author} {\bibinfo {author} {\bibfnamefont {M.}~\bibnamefont
  {Barnes}}, \bibinfo {author} {\bibfnamefont {F.~I.}\ \bibnamefont {Parra}}, \
  and\ \bibinfo {author} {\bibfnamefont {A.~A.}\ \bibnamefont {Schekochihin}},\
  }\href {\doibase 10.1103/PhysRevLett.107.115003} {\bibfield  {journal}
  {\bibinfo  {journal} {Phys. Rev. Lett.}\ }\textbf {\bibinfo {volume} {107}},\
  \bibinfo {pages} {115003} (\bibinfo {year} {2011})}\BibitemShut {NoStop}%
\bibitem [{\citenamefont {Nakata}, \citenamefont {Watanabe},\ and\
  \citenamefont {Sugama}(2012)}]{Nakata:2012p1387}%
  \BibitemOpen
  \bibfield  {author} {\bibinfo {author} {\bibfnamefont {M.}~\bibnamefont
  {Nakata}}, \bibinfo {author} {\bibfnamefont {T.-H.}\ \bibnamefont
  {Watanabe}}, \ and\ \bibinfo {author} {\bibfnamefont {H.}~\bibnamefont
  {Sugama}},\ }\href {\doibase 10.1063/1.3675855} {\bibfield  {journal}
  {\bibinfo  {journal} {Phys. Plasmas}\ }\textbf {\bibinfo {volume} {19}},\
  \bibinfo {pages} {2303} (\bibinfo {year} {2012})}\BibitemShut {NoStop}%
\bibitem [{\citenamefont {Morel}\ \emph {et~al.}(2011)\citenamefont {Morel},
  \citenamefont {Navarro}, \citenamefont {Albrecht-Marc}, \citenamefont
  {Carati}, \citenamefont {Merz}, \citenamefont {G{\"o}rler},\ and\
  \citenamefont {Jenko}}]{Morel:2011p1339}%
  \BibitemOpen
  \bibfield  {author} {\bibinfo {author} {\bibfnamefont {P.}~\bibnamefont
  {Morel}}, \bibinfo {author} {\bibfnamefont {A.~B.}\ \bibnamefont {Navarro}},
  \bibinfo {author} {\bibfnamefont {M.}~\bibnamefont {Albrecht-Marc}}, \bibinfo
  {author} {\bibfnamefont {D.}~\bibnamefont {Carati}}, \bibinfo {author}
  {\bibfnamefont {F.}~\bibnamefont {Merz}}, \bibinfo {author} {\bibfnamefont
  {T.}~\bibnamefont {G{\"o}rler}}, \ and\ \bibinfo {author} {\bibfnamefont
  {F.}~\bibnamefont {Jenko}},\ }\href {\doibase 10.1063/1.3601053} {\bibfield
  {journal} {\bibinfo  {journal} {Phys. Plasmas}\ }\textbf {\bibinfo {volume}
  {18}},\ \bibinfo {pages} {2301} (\bibinfo {year} {2011})}\BibitemShut
  {NoStop}%
\bibitem [{\citenamefont {Morel}\ \emph {et~al.}(2012)\citenamefont {Morel},
  \citenamefont {Navarro}, \citenamefont {Albrecht-Marc}, \citenamefont
  {Carati}, \citenamefont {Merz}, \citenamefont {G{\"o}rler},\ and\
  \citenamefont {Jenko}}]{Morel:2012p1390}%
  \BibitemOpen
  \bibfield  {author} {\bibinfo {author} {\bibfnamefont {P.}~\bibnamefont
  {Morel}}, \bibinfo {author} {\bibfnamefont {A.~B.}\ \bibnamefont {Navarro}},
  \bibinfo {author} {\bibfnamefont {M.}~\bibnamefont {Albrecht-Marc}}, \bibinfo
  {author} {\bibfnamefont {D.}~\bibnamefont {Carati}}, \bibinfo {author}
  {\bibfnamefont {F.}~\bibnamefont {Merz}}, \bibinfo {author} {\bibfnamefont
  {T.}~\bibnamefont {G{\"o}rler}}, \ and\ \bibinfo {author} {\bibfnamefont
  {F.}~\bibnamefont {Jenko}},\ }\href {\doibase 10.1063/1.3677366} {\bibfield
  {journal} {\bibinfo  {journal} {Phys. Plasmas}\ }\textbf {\bibinfo {volume}
  {19}},\ \bibinfo {pages} {2311} (\bibinfo {year} {2012})}\BibitemShut
  {NoStop}%
\bibitem [{\citenamefont {Navarro}\ \emph {et~al.}(2014)\citenamefont
  {Navarro}, \citenamefont {Teaca}, \citenamefont {Jenko}, \citenamefont
  {Hammett},\ and\ \citenamefont {Happel}}]{BanonNavarro:2014p1535}%
  \BibitemOpen
  \bibfield  {author} {\bibinfo {author} {\bibfnamefont {A.~B.}\ \bibnamefont
  {Navarro}}, \bibinfo {author} {\bibfnamefont {B.}~\bibnamefont {Teaca}},
  \bibinfo {author} {\bibfnamefont {F.}~\bibnamefont {Jenko}}, \bibinfo
  {author} {\bibfnamefont {G.~W.}\ \bibnamefont {Hammett}}, \ and\ \bibinfo
  {author} {\bibfnamefont {T.}~\bibnamefont {Happel}},\ }\href {\doibase
  10.1063/1.4868235} {\bibfield  {journal} {\bibinfo  {journal} {Phys.
  Plasmas}\ }\textbf {\bibinfo {volume} {21}},\ \bibinfo {pages} {032304}
  (\bibinfo {year} {2014})}\BibitemShut {NoStop}%
\bibitem [{\citenamefont {Kraichnan}(1959)}]{Kraichnan:1959p497}%
  \BibitemOpen
  \bibfield  {author} {\bibinfo {author} {\bibfnamefont {R.~H.}\ \bibnamefont
  {Kraichnan}},\ }\href@noop {} {\bibfield  {journal} {\bibinfo  {journal} {J.
  Fluid Mech.}\ }\textbf {\bibinfo {volume} {5}},\ \bibinfo {pages} {497}
  (\bibinfo {year} {1959})}\BibitemShut {NoStop}%
\bibitem [{\citenamefont {Teaca}, \citenamefont {Carati},\ and\ \citenamefont
  {Domaradzki}(2011)}]{Teaca:2011p1362}%
  \BibitemOpen
  \bibfield  {author} {\bibinfo {author} {\bibfnamefont {B.}~\bibnamefont
  {Teaca}}, \bibinfo {author} {\bibfnamefont {D.}~\bibnamefont {Carati}}, \
  and\ \bibinfo {author} {\bibfnamefont {J.~A.}\ \bibnamefont {Domaradzki}},\
  }\href {\doibase 10.1063/1.3661086} {\bibfield  {journal} {\bibinfo
  {journal} {Phys. Plasmas}\ }\textbf {\bibinfo {volume} {18}},\ \bibinfo
  {pages} {2307} (\bibinfo {year} {2011})}\BibitemShut {NoStop}%
\bibitem [{\citenamefont {Beer}, \citenamefont {Cowley},\ and\ \citenamefont
  {Hammett}(1995)}]{Beer:1995p1292}%
  \BibitemOpen
  \bibfield  {author} {\bibinfo {author} {\bibfnamefont {M.~A.}\ \bibnamefont
  {Beer}}, \bibinfo {author} {\bibfnamefont {S.~C.}\ \bibnamefont {Cowley}}, \
  and\ \bibinfo {author} {\bibfnamefont {G.~W.}\ \bibnamefont {Hammett}},\
  }\href {\doibase 10.1063/1.871232} {\bibfield  {journal} {\bibinfo  {journal}
  {Phys. Plasmas}\ }\textbf {\bibinfo {volume} {2}},\ \bibinfo {pages} {2687}
  (\bibinfo {year} {1995})}\BibitemShut {NoStop}%
\bibitem [{\citenamefont {Brizard}(1989)}]{Brizard:1989p1227}%
  \BibitemOpen
  \bibfield  {author} {\bibinfo {author} {\bibfnamefont {A.}~\bibnamefont
  {Brizard}},\ }\href {\doibase 10.1063/1.858968} {\bibfield  {journal}
  {\bibinfo  {journal} {Physics of Fluids B}\ }\textbf {\bibinfo {volume}
  {1}},\ \bibinfo {pages} {1381} (\bibinfo {year} {1989})}\BibitemShut
  {NoStop}%
\bibitem [{\citenamefont {Hammett}\ \emph {et~al.}(1993)\citenamefont
  {Hammett}, \citenamefont {Beer}, \citenamefont {Dorland}, \citenamefont
  {Cowley},\ and\ \citenamefont {Smith}}]{Hammett:1993p1536}%
  \BibitemOpen
  \bibfield  {author} {\bibinfo {author} {\bibfnamefont {G.~W.}\ \bibnamefont
  {Hammett}}, \bibinfo {author} {\bibfnamefont {M.~A.}\ \bibnamefont {Beer}},
  \bibinfo {author} {\bibfnamefont {W.}~\bibnamefont {Dorland}}, \bibinfo
  {author} {\bibfnamefont {S.~C.}\ \bibnamefont {Cowley}}, \ and\ \bibinfo
  {author} {\bibfnamefont {S.~A.}\ \bibnamefont {Smith}},\ }\href {\doibase
  10.1088/0741-3335/35/8/006} {\bibfield  {journal} {\bibinfo  {journal}
  {Plasma Phys. Control. Fusion}\ }\textbf {\bibinfo {volume} {35}},\ \bibinfo
  {pages} {973} (\bibinfo {year} {1993})}\BibitemShut {NoStop}%
%
\bibitem{Merz2009} F. Merz, Ph.D. thesis, Universit\"{a}t M\"{u}nster, (2009).
 %
\bibitem [{\citenamefont {Jenko}\ \emph {et~al.}(2000)\citenamefont {Jenko},
  \citenamefont {Dorland}, \citenamefont {Kotschenreuther},\ and\ \citenamefont
  {Rogers}}]{Jenko:2000p1248}%
  \BibitemOpen
  \bibfield  {author} {\bibinfo {author} {\bibfnamefont {F.}~\bibnamefont
  {Jenko}}, \bibinfo {author} {\bibfnamefont {W.}~\bibnamefont {Dorland}},
  \bibinfo {author} {\bibfnamefont {M.}~\bibnamefont {Kotschenreuther}}, \ and\
  \bibinfo {author} {\bibfnamefont {B.~N.}\ \bibnamefont {Rogers}},\ }\href
  {\doibase 10.1063/1.874014} {\bibfield  {journal} {\bibinfo  {journal} {Phys.
  Plasmas}\ }\textbf {\bibinfo {volume} {7}},\ \bibinfo {pages} {1904}
  (\bibinfo {year} {2000})}\BibitemShut {NoStop}%
\bibitem [{\citenamefont {G{\"o}rler}\ \emph {et~al.}(2011)\citenamefont
  {G{\"o}rler}, \citenamefont {Lapillonne}, \citenamefont {Brunner},
  \citenamefont {Dannert}, \citenamefont {Jenko}, \citenamefont {Merz},\ and\
  \citenamefont {Told}}]{Gorler:2011p1340}%
  \BibitemOpen
  \bibfield  {author} {\bibinfo {author} {\bibfnamefont {T.}~\bibnamefont
  {G{\"o}rler}}, \bibinfo {author} {\bibfnamefont {X.}~\bibnamefont
  {Lapillonne}}, \bibinfo {author} {\bibfnamefont {S.}~\bibnamefont {Brunner}},
  \bibinfo {author} {\bibfnamefont {T.}~\bibnamefont {Dannert}}, \bibinfo
  {author} {\bibfnamefont {F.}~\bibnamefont {Jenko}}, \bibinfo {author}
  {\bibfnamefont {F.}~\bibnamefont {Merz}}, \ and\ \bibinfo {author}
  {\bibfnamefont {D.}~\bibnamefont {Told}},\ }\href {\doibase
  10.1016/j.jcp.2011.05.034} {\bibfield  {journal} {\bibinfo  {journal}
  {Journal of Computational Physics}\ }\textbf {\bibinfo {volume} {230}},\
  \bibinfo {pages} {7053} (\bibinfo {year} {2011})}\BibitemShut {NoStop}%
\bibitem [{\citenamefont {Lapillonne}\ \emph {et~al.}(2009)\citenamefont
  {Lapillonne}, \citenamefont {Brunner}, \citenamefont {Dannert}, \citenamefont
  {Jolliet}, \citenamefont {Marinoni}, \citenamefont {Villard}, \citenamefont
  {G{\"o}rler}, \citenamefont {Jenko},\ and\ \citenamefont
  {Merz}}]{Lapillonne:2009p1355}%
  \BibitemOpen
  \bibfield  {author} {\bibinfo {author} {\bibfnamefont {X.}~\bibnamefont
  {Lapillonne}}, \bibinfo {author} {\bibfnamefont {S.}~\bibnamefont {Brunner}},
  \bibinfo {author} {\bibfnamefont {T.}~\bibnamefont {Dannert}}, \bibinfo
  {author} {\bibfnamefont {S.}~\bibnamefont {Jolliet}}, \bibinfo {author}
  {\bibfnamefont {A.}~\bibnamefont {Marinoni}}, \bibinfo {author}
  {\bibfnamefont {L.}~\bibnamefont {Villard}}, \bibinfo {author} {\bibfnamefont
  {T.}~\bibnamefont {G{\"o}rler}}, \bibinfo {author} {\bibfnamefont
  {F.}~\bibnamefont {Jenko}}, \ and\ \bibinfo {author} {\bibfnamefont
  {F.}~\bibnamefont {Merz}},\ }\href {\doibase 10.1063/1.3096710} {\bibfield
  {journal} {\bibinfo  {journal} {Phys. Plasmas}\ }\textbf {\bibinfo {volume}
  {16}},\ \bibinfo {pages} {2308} (\bibinfo {year} {2009})}\BibitemShut
  {NoStop}%
\bibitem [{\citenamefont {Dimits}\ \emph {et~al.}(2000)\citenamefont {Dimits},
  \citenamefont {Bateman}, \citenamefont {Beer}, \citenamefont {Cohen},
  \citenamefont {Dorland}, \citenamefont {Hammett}, \citenamefont {Kim},
  \citenamefont {Kinsey}, \citenamefont {Kotschenreuther}, \citenamefont
  {Kritz}, \citenamefont {Lao}, \citenamefont {Mandrekas}, \citenamefont
  {Nevins}, \citenamefont {Parker}, \citenamefont {Redd}, \citenamefont
  {Shumaker}, \citenamefont {Sydora},\ and\ \citenamefont
  {Weiland}}]{Dimits:2000p1375}%
  \BibitemOpen
  \bibfield  {author} {\bibinfo {author} {\bibfnamefont {A.~M.}\ \bibnamefont
  {Dimits}}, \bibinfo {author} {\bibfnamefont {G.}~\bibnamefont {Bateman}},
  \bibinfo {author} {\bibfnamefont {M.~A.}\ \bibnamefont {Beer}}, \bibinfo
  {author} {\bibfnamefont {B.~I.}\ \bibnamefont {Cohen}}, \bibinfo {author}
  {\bibfnamefont {W.}~\bibnamefont {Dorland}}, \bibinfo {author} {\bibfnamefont
  {G.~W.}\ \bibnamefont {Hammett}}, \bibinfo {author} {\bibfnamefont
  {C.}~\bibnamefont {Kim}}, \bibinfo {author} {\bibfnamefont {J.~E.}\
  \bibnamefont {Kinsey}}, \bibinfo {author} {\bibfnamefont {M.}~\bibnamefont
  {Kotschenreuther}}, \bibinfo {author} {\bibfnamefont {A.~H.}\ \bibnamefont
  {Kritz}}, \bibinfo {author} {\bibfnamefont {L.~L.}\ \bibnamefont {Lao}},
  \bibinfo {author} {\bibfnamefont {J.}~\bibnamefont {Mandrekas}}, \bibinfo
  {author} {\bibfnamefont {W.~M.}\ \bibnamefont {Nevins}}, \bibinfo {author}
  {\bibfnamefont {S.~E.}\ \bibnamefont {Parker}}, \bibinfo {author}
  {\bibfnamefont {A.~J.}\ \bibnamefont {Redd}}, \bibinfo {author}
  {\bibfnamefont {D.~E.}\ \bibnamefont {Shumaker}}, \bibinfo {author}
  {\bibfnamefont {R.}~\bibnamefont {Sydora}}, \ and\ \bibinfo {author}
  {\bibfnamefont {J.}~\bibnamefont {Weiland}},\ }\href {\doibase
  10.1063/1.873896} {\bibfield  {journal} {\bibinfo  {journal} {Phys. Plasmas}\
  }\textbf {\bibinfo {volume} {7}},\ \bibinfo {pages} {969} (\bibinfo {year}
  {2000})}\BibitemShut {NoStop}%
\bibitem [{\citenamefont {Nevins}\ \emph {et~al.}(2006)\citenamefont {Nevins},
  \citenamefont {Candy}, \citenamefont {Cowley}, \citenamefont {Dannert},
  \citenamefont {Dimits}, \citenamefont {Dorland}, \citenamefont
  {Estrada-Mila}, \citenamefont {Hammett}, \citenamefont {Jenko}, \citenamefont
  {Pueschel},\ and\ \citenamefont {Shumaker}}]{Nevins:2006p1454}%
  \BibitemOpen
  \bibfield  {author} {\bibinfo {author} {\bibfnamefont {W.~M.}\ \bibnamefont
  {Nevins}}, \bibinfo {author} {\bibfnamefont {J.}~\bibnamefont {Candy}},
  \bibinfo {author} {\bibfnamefont {S.}~\bibnamefont {Cowley}}, \bibinfo
  {author} {\bibfnamefont {T.}~\bibnamefont {Dannert}}, \bibinfo {author}
  {\bibfnamefont {A.}~\bibnamefont {Dimits}}, \bibinfo {author} {\bibfnamefont
  {W.}~\bibnamefont {Dorland}}, \bibinfo {author} {\bibfnamefont
  {C.}~\bibnamefont {Estrada-Mila}}, \bibinfo {author} {\bibfnamefont {G.~W.}\
  \bibnamefont {Hammett}}, \bibinfo {author} {\bibfnamefont {F.}~\bibnamefont
  {Jenko}}, \bibinfo {author} {\bibfnamefont {M.~J.}\ \bibnamefont {Pueschel}},
  \ and\ \bibinfo {author} {\bibfnamefont {D.~E.}\ \bibnamefont {Shumaker}},\
  }\href {\doibase 10.1063/1.2402510} {\bibfield  {journal} {\bibinfo
  {journal} {Phys. Plasmas}\ }\textbf {\bibinfo {volume} {13}},\ \bibinfo
  {pages} {2306} (\bibinfo {year} {2006})}\BibitemShut {NoStop}%
\bibitem [{\citenamefont {Dorland}\ \emph {et~al.}(2000)\citenamefont
  {Dorland}, \citenamefont {Jenko}, \citenamefont {Kotschenreuther},\ and\
  \citenamefont {Rogers}}]{Dorland:2000p1088}%
  \BibitemOpen
  \bibfield  {author} {\bibinfo {author} {\bibfnamefont {W.}~\bibnamefont
  {Dorland}}, \bibinfo {author} {\bibfnamefont {F.}~\bibnamefont {Jenko}},
  \bibinfo {author} {\bibfnamefont {M.}~\bibnamefont {Kotschenreuther}}, \ and\
  \bibinfo {author} {\bibfnamefont {B.~N.}\ \bibnamefont {Rogers}},\ }\href
  {\doibase 10.1103/PhysRevLett.85.5579} {\bibfield  {journal} {\bibinfo
  {journal} {Phys. Rev. Lett.}\ }\textbf {\bibinfo {volume} {85}},\ \bibinfo
  {pages} {5579} (\bibinfo {year} {2000})}\BibitemShut {NoStop}%
\bibitem [{\citenamefont {Schekochihin}\ \emph {et~al.}(2009)\citenamefont
  {Schekochihin}, \citenamefont {Cowley}, \citenamefont {Dorland},
  \citenamefont {Hammett}, \citenamefont {Howes}, \citenamefont {Quataert},\
  and\ \citenamefont {Tatsuno}}]{Schekochihin:2009p1131}%
  \BibitemOpen
  \bibfield  {author} {\bibinfo {author} {\bibfnamefont {A.~A.}\ \bibnamefont
  {Schekochihin}}, \bibinfo {author} {\bibfnamefont {S.~C.}\ \bibnamefont
  {Cowley}}, \bibinfo {author} {\bibfnamefont {W.}~\bibnamefont {Dorland}},
  \bibinfo {author} {\bibfnamefont {G.~W.}\ \bibnamefont {Hammett}}, \bibinfo
  {author} {\bibfnamefont {G.~G.}\ \bibnamefont {Howes}}, \bibinfo {author}
  {\bibfnamefont {E.}~\bibnamefont {Quataert}}, \ and\ \bibinfo {author}
  {\bibfnamefont {T.}~\bibnamefont {Tatsuno}},\ }\href {\doibase
  10.1088/0067-0049/182/1/310} {\bibfield  {journal} {\bibinfo  {journal} {The
  Astrophysical Journal Supplement}\ }\textbf {\bibinfo {volume} {182}},\
  \bibinfo {pages} {310} (\bibinfo {year} {2009})}\BibitemShut {NoStop}%
    %
\bibitem{ABNThesis} A. {Ba\~n\'on Navarro}, Ph.D. thesis, Universit\'e libre de Bruxelles, (2012).
 %
  \bibitem [{\citenamefont {Verma}(2004)}]{Verma:2004p206}%
  \BibitemOpen
  \bibfield  {author} {\bibinfo {author} {\bibfnamefont {M.~K.}\ \bibnamefont
  {Verma}},\ }\href {\doibase 10.1016/j.physrep.2004.07.007} {\bibfield
  {journal} {\bibinfo  {journal} {Physics Reports}\ }\textbf {\bibinfo {volume}
  {401}},\ \bibinfo {pages} {229} (\bibinfo {year} {2004})}\BibitemShut
  {NoStop}%
  %
\bibitem{Howes:2011p1459}  G. G. Howes, J. M. Tenbarge and W. Dorland,{Phys. Plasmas} {\bf 18}, {2305} ({2011}). 
 %
\bibitem [{\citenamefont {Navarro}\ \emph {et~al.}(2011)\citenamefont
  {Navarro}, \citenamefont {Morel}, \citenamefont {Albrecht-Marc},
  \citenamefont {Carati}, \citenamefont {Merz}, \citenamefont {G{\"o}rler},\
  and\ \citenamefont {Jenko}}]{BanonNavarro:2011p1274}%
  \BibitemOpen
  \bibfield  {author} {\bibinfo {author} {\bibfnamefont {A.~B.}\ \bibnamefont
  {Navarro}}, \bibinfo {author} {\bibfnamefont {P.}~\bibnamefont {Morel}},
  \bibinfo {author} {\bibfnamefont {M.}~\bibnamefont {Albrecht-Marc}}, \bibinfo
  {author} {\bibfnamefont {D.}~\bibnamefont {Carati}}, \bibinfo {author}
  {\bibfnamefont {F.}~\bibnamefont {Merz}}, \bibinfo {author} {\bibfnamefont
  {T.}~\bibnamefont {G{\"o}rler}}, \ and\ \bibinfo {author} {\bibfnamefont
  {F.}~\bibnamefont {Jenko}},\ }\href {\doibase 10.1103/PhysRevLett.106.055001}
  {\bibfield  {journal} {\bibinfo  {journal} {Phys. Rev. Lett.}\ }\textbf
  {\bibinfo {volume} {106}},\ \bibinfo {pages} {55001} (\bibinfo {year}
  {2011})}\BibitemShut {NoStop}%
\bibitem [{\citenamefont {Teaca}\ \emph {et~al.}(2012)\citenamefont {Teaca},
  \citenamefont {Navarro}, \citenamefont {Jenko}, \citenamefont {Brunner},\
  and\ \citenamefont {Villard}}]{Teaca:2012p1415}%
  \BibitemOpen
  \bibfield  {author} {\bibinfo {author} {\bibfnamefont {B.}~\bibnamefont
  {Teaca}}, \bibinfo {author} {\bibfnamefont {A.~B.}\ \bibnamefont {Navarro}},
  \bibinfo {author} {\bibfnamefont {F.}~\bibnamefont {Jenko}}, \bibinfo
  {author} {\bibfnamefont {S.}~\bibnamefont {Brunner}}, \ and\ \bibinfo
  {author} {\bibfnamefont {L.}~\bibnamefont {Villard}},\ }\href {\doibase
  10.1103/PhysRevLett.109.235003} {\bibfield  {journal} {\bibinfo  {journal}
  {Phys. Rev. Lett.}\ }\textbf {\bibinfo {volume} {109}},\ \bibinfo {pages}
  {235003} (\bibinfo {year} {2012})}\BibitemShut {NoStop}%
\bibitem [{\citenamefont {Tatsuno}\ \emph {et~al.}(2010)\citenamefont
  {Tatsuno}, \citenamefont {Barnes}, \citenamefont {Cowley}, \citenamefont
  {Dorland}, \citenamefont {Howes}, \citenamefont {Numata}, \citenamefont
  {Plunk},\ and\ \citenamefont {Schekochihin}}]{Tatsuno:2010p1363}%
  \BibitemOpen
  \bibfield  {author} {\bibinfo {author} {\bibfnamefont {T.}~\bibnamefont
  {Tatsuno}}, \bibinfo {author} {\bibfnamefont {M.}~\bibnamefont {Barnes}},
  \bibinfo {author} {\bibfnamefont {S.}~\bibnamefont {Cowley}}, \bibinfo
  {author} {\bibfnamefont {W.}~\bibnamefont {Dorland}}, \bibinfo {author}
  {\bibfnamefont {G.}~\bibnamefont {Howes}}, \bibinfo {author} {\bibfnamefont
  {R.}~\bibnamefont {Numata}}, \bibinfo {author} {\bibfnamefont
  {G.}~\bibnamefont {Plunk}}, \ and\ \bibinfo {author} {\bibfnamefont
  {A.}~\bibnamefont {Schekochihin}},\ }\href@noop {} {\bibfield  {journal}
  {\bibinfo  {journal} {J. Plasma Fusion Res}\ }\textbf {\bibinfo {volume}
  {9}},\ \bibinfo {pages} {509} (\bibinfo {year} {2010})}\BibitemShut {NoStop}%
 %
\bibitem{Camargo:1995p1564}  S. J. Camargo, D. Biskamp, B. D. Scott,{ Phys. Plasmas} {\bf 2}, {48} ({1995}). 
 %
 \bibitem{Domaradzki:1990p145}  A. J. Domaradzki and R. S. Rogallo,{ Phys. Fluids} {\bf 2}, {413} ({1990}). 
 %
  \bibitem [{\citenamefont {Debliquy}, \citenamefont {Verma},\ and\ \citenamefont
  {Carati}(2005)}]{Debliquy:2005p203}%
  \BibitemOpen
  \bibfield  {author} {\bibinfo {author} {\bibfnamefont {O.}~\bibnamefont
  {Debliquy}}, \bibinfo {author} {\bibfnamefont {M.~K.}\ \bibnamefont {Verma}},
  \ and\ \bibinfo {author} {\bibfnamefont {D.}~\bibnamefont {Carati}},\ }\href
  {\doibase 10.1063/1.1867996} {\bibfield  {journal} {\bibinfo  {journal}
  {Phys. Plasmas}\ }\textbf {\bibinfo {volume} {12}},\ \bibinfo {pages} {2309}
  (\bibinfo {year} {2005})}\BibitemShut {NoStop}%
\bibitem [{\citenamefont {Eyink}(2005)}]{Eyink:2005p854}%
  \BibitemOpen
  \bibfield  {author} {\bibinfo {author} {\bibfnamefont {G.~L.}\ \bibnamefont
  {Eyink}},\ }\href {\doibase 10.1016/j.physd.2005.05.018} {\bibfield
  {journal} {\bibinfo  {journal} {Physica D}\ }\textbf {\bibinfo {volume}
  {207}},\ \bibinfo {pages} {91} (\bibinfo {year} {2005})}\BibitemShut
  {NoStop}%
\bibitem [{\citenamefont {Domaradzki}, \citenamefont {Teaca},\ and\
  \citenamefont {Carati}(2009)}]{Domaradzki:2009p557}%
  \BibitemOpen
  \bibfield  {author} {\bibinfo {author} {\bibfnamefont {J.~A.}\ \bibnamefont
  {Domaradzki}}, \bibinfo {author} {\bibfnamefont {B.}~\bibnamefont {Teaca}}, \
  and\ \bibinfo {author} {\bibfnamefont {D.}~\bibnamefont {Carati}},\ }\href
  {\doibase 10.1063/1.3081558} {\bibfield  {journal} {\bibinfo  {journal}
  {Phys. Fluids}\ }\textbf {\bibinfo {volume} {21}},\ \bibinfo {pages} {5106}
  (\bibinfo {year} {2009})}\BibitemShut {NoStop}%
\bibitem [{\citenamefont {Domaradzki}, \citenamefont {Teaca},\ and\
  \citenamefont {Carati}(2010)}]{Domaradzki:2010p1012}%
  \BibitemOpen
  \bibfield  {author} {\bibinfo {author} {\bibfnamefont {J.~A.}\ \bibnamefont
  {Domaradzki}}, \bibinfo {author} {\bibfnamefont {B.}~\bibnamefont {Teaca}}, \
  and\ \bibinfo {author} {\bibfnamefont {D.}~\bibnamefont {Carati}},\ }\href
  {\doibase 10.1063/1.3431227} {\bibfield  {journal} {\bibinfo  {journal}
  {Phys. Fluids}\ }\textbf {\bibinfo {volume} {22}},\ \bibinfo {pages} {1702}
  (\bibinfo {year} {2010})}\BibitemShut {NoStop}%
\bibitem [{\citenamefont {Zhou}(1993)}]{Zhou:1993p949}%
  \BibitemOpen
  \bibfield  {author} {\bibinfo {author} {\bibfnamefont {Y.}~\bibnamefont
  {Zhou}},\ }\href {\doibase 10.1063/1.858593} {\bibfield  {journal} {\bibinfo
  {journal} {Phys. Fluids}\ }\textbf {\bibinfo {volume} {5}},\ \bibinfo {pages}
  {1092} (\bibinfo {year} {1993})}\BibitemShut {NoStop}%
\bibitem [{\citenamefont {Eyink}\ and\ \citenamefont
  {Aluie}(2009)}]{Eyink:2009p808}%
  \BibitemOpen
  \bibfield  {author} {\bibinfo {author} {\bibfnamefont {G.~L.}\ \bibnamefont
  {Eyink}}\ and\ \bibinfo {author} {\bibfnamefont {H.}~\bibnamefont {Aluie}},\
  }\href {\doibase 10.1063/1.3266883} {\bibfield  {journal} {\bibinfo
  {journal} {Phys. Fluids}\ }\textbf {\bibinfo {volume} {21}},\ \bibinfo
  {pages} {5107} (\bibinfo {year} {2009})},\ \bibinfo {note} {(c) 2009:
  American Institute of Physics}\BibitemShut {NoStop}%
\bibitem [{\citenamefont {Aluie}\ and\ \citenamefont
  {Eyink}(2009)}]{Aluie:2009p809}%
  \BibitemOpen
  \bibfield  {author} {\bibinfo {author} {\bibfnamefont {H.}~\bibnamefont
  {Aluie}}\ and\ \bibinfo {author} {\bibfnamefont {G.~L.}\ \bibnamefont
  {Eyink}},\ }\href {\doibase 10.1063/1.3266948} {\bibfield  {journal}
  {\bibinfo  {journal} {Phys. Fluids}\ }\textbf {\bibinfo {volume} {21}},\
  \bibinfo {pages} {5108} (\bibinfo {year} {2009})},\ \bibinfo {note} {(c)
  2009: American Institute of Physics}\BibitemShut {NoStop}%
\end{thebibliography}

%

\end{document}